\documentclass[a4paper,11pt]{article}
\pdfoutput=1

\usepackage{latex/jinstpub} 
\usepackage[binary-units=true]{siunitx}
\usepackage[sicmds,freestanding]{latex/hepunits}
\usepackage{latex/uk-pre-ms-tb-paper-defs}
\usepackage{lineno}
\usepackage{xspace}
\usepackage{multirow}
\usepackage{booktabs}
\usepackage{caption}
\usepackage{subcaption}
\usepackage{comment}

\title{\boldmath Testbeam analysis of biasing structures for irradiated hybrid pixel detectors}

\author[a,1,2]{A. G. Rennie \note{Corresponding author.}\note{Previously at University of Glasgow, School of Physics and Astronomy, Kelvin Building, University Avenue, Glasgow, G12 8QQ.}}
\author[b]{C. M.  Buttar}
\author[c]{Y. Gao}
\author[d]{R. Gonz{\'a}lez L{\'o}pez}
\author[b]{D. Maneuski}
\author[c]{E. Pender}
\author[e]{~~~Q. Qin}
\author[d]{M. Sullivan}
\author[d]{J. T. Taylor}
\author[b]{K. Wraight}

\affiliation[a]{University of California, Irvine\\Department of Physics and Astronomy, 4129 Frederick Reines Hall, Irvine, CA 92697-4575}
\affiliation[b]{University of Glasgow\\School of Physics and Astronomy, Kelvin Building, University Avenue, Glasgow, G12 8QQ}
\affiliation[c]{University of Edinburgh\\School of Physics and Astronomy, James Clerk Maxwell Building, Peter Guthrie Tait Road, Edinburgh, EH9 3FD }
\affiliation[d]{University of Liverpool\\Department of Physics, Oliver Lodge,	Oxford Street, Liverpool, L69 7ZE}
\affiliation[e]{University of Manchester\\Department of Physics and Astronomy, Schuster Building, Oxford Road, Manchester, M13 9PL}

\emailAdd{adam.rennie@cern.ch}

\abstract{Following the Phase-II upgrade during Long Shutdown~(LS3), the LHC aims to reach a peak
	instantaneous luminosity of $7.5\times 10^{34}$~\invcmsqpersec, which
	corresponds to an average of around 200 inelastic proton-proton collisions per beam-crossing~(every \SI{25}{\ns}).
	To cope with these conditions, the ATLAS Inner Detector will be replaced
	by a new all-silicon system --- the Inner Tracker~(ITk).
	The ITk will be operational for more than ten years, during
	which time ATLAS is expected to record approximately 4000~\invfb~of data.
	The ITk's pixel sub-system is based on hybrid pixel modules with
	new silicon sensors and readout chips. 
	These studies focus on testbeam campaigns undertaken to study
	the spatial resolution and efficiencies of hybrid pixel detector modules based on the first large-structure prototype front-end readout chip --- the RD53A ---  using planar silicon sensors. These devices have been irradiated to replicate the effect of the high radiation environment present during operation in the ATLAS detector.
	Results for devices using sensors with different punch-through bias structures and using different readout modes are summarised. Those with sensors incorporating a punch-through bias structure are found to exhibit systematically lower efficiency than those without, as a result of local areas of relative inefficiency around the punch-through dots. Despite this, all devices measured are found to satisfy the requirement of 97\% efficiency at $\bias=\SI{400}{\volt}$ after being irradiated to end-of-life fluence.}

\keywords{Hybrid detectors, Radiation-hard detectors, Particle tracking detectors (Solid-state detectors), Performance of High Energy Physics Detectors}

\begin{document}
\maketitle
\flushbottom

\newpage
\section{Introduction}\label{sec:introduction}

For the High Luminosity era of the Large Hadron Collider~(HL-LHC),
the present ATLAS Inner Detector will be replaced by a new all-silicon Inner Tracker~(ITk)~\cite{macchiolo_phase-2_2020,keller_atlas_2020}
in order to cope with increased occupancy and radiation.
The ITk will consist of a pixel detector closest to the beamline and a strip detector for the outer section. This will provide coverage for
charged particle reconstruction up to~$|\eta|<4$. 
The ATLAS upgrade physics programme drives the design and performance requirements of the pixel detector. 
The demand for high precision, radiation hard, rapid readout pixel modules has required the design 
of a new front-end readout chip and sensor architecture to 
meet the performance requirements necessary for the HL-LHC environment.

The pixel detector will consist of five barrel layers in the central region,
and a number of ring-shaped  layers in the forward region, leading to a total active area of around \SI{13}{\metre\squared}.
The innermost two layers of the ITk pixel detector~\cite{M_bius_2022} --- the \emph{inner system} --- will experience the highest total ionising dose and so will feature 3D silicon sensors, which have heightened radiation hardness. The inner system is designed to be replaced after \SI{2000}{\invfb}.
All remaining layers --- the \emph{outer system} --- will be based on planar silicon sensors with thickness \SI{150}{\um}. Novel front-end ASICs, implemented in \SI{65}{\nano\metre} technology,
are connected to the silicon sensors using bump-bonding
to form a bare modules. This is then glued and wire-bonded to a flexible printed circuit board~(PCB). 
The off-detector readout electronics will be implemented in the framework of the general ATLAS trigger and
DAQ system with a readout rate of up to
5~Gb/s per data link for the innermost layers.

The outer system is expected to experience a fluence up to \SI{5e15}{\nequ\per\cm\squared} and a total ionising dose of \SI{5}{\mega\gray}.

Testbeam measurements are vital to study, understand,
and verify the performance of the new readout chips and sensor technologies.
This paper summarises several testbeam campaigns undertaken for
several R\&D sensors developed by the ATLAS UK ITk community. 
Different biasing structures, readout modes, and pixel module operation
parameters such as the bias voltage and threshold are studied in detail using irradiated devices in order to mimic the effects of LHC operation on the detector modules.

\section{Devices under test}\label{sec:rd53a}

The devices under test~(DUTs) are hybrid pixel modules, 
each including a passive high resistivity silicon sensor~(\textit{n-in-p})
and a front-end readout chip combined by flip-chip bump-bonding, and a flexible PCB.
In this paper, prototype modules with different silicon sensor designs, together with
a prototype front-end readout chip, the RD53A~\cite{Garcia-Sciveres:2287593}, are characterised. The sensors were manufactured by Micron Semiconductor Ltd.

The DUTs presented here were all irradiated at the Karlsruher Institut für Technologie~(KIT) to $3.4\times10^{15}$~\SI{}{\nequ\per\cm\squared} using \SI{25}{\MeV} protons extracted from the Karlsruhe Kompakt Zyklotron~\cite{KIT}. This fluence is based on the expected radiation levels in the outer system of the ATLAS pixel detector.

\begin{figure}[btp]
	\centering
	\includegraphics[width=\textwidth, trim=0cm 4cm 0cm 0cm, clip]{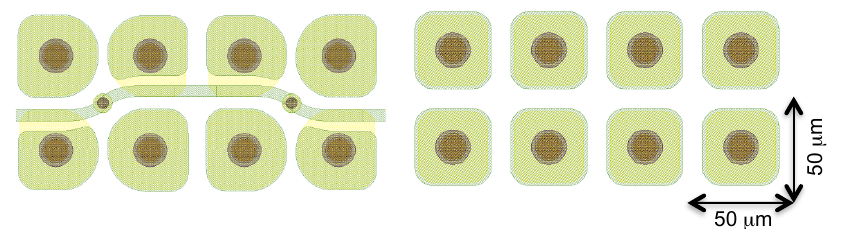}
	\caption{Pixel sensor designs with~(left) and without~(right) biasing structures~\cite{ATLAS-TDR-30}. The design with biasing structure implemented uses punch-through dots common to four pixel cells, connected to a biasing rail in a zigzag configuration.}
	\label{fig:PTB}
\end{figure}

Devices made from two different planar silicon $n$-in-$p$ sensors are presented. Both sensors have the same pixel pitch~($50\times50$~\SI{}{\um\squared}) and thickness~(\SI{150}{\um}), but are differentiated by \emph{punch-through bias}~(PTB) structure. Once in operation, the PTB structure is inactive. However, it serves a vital role in the production of the pixel modules in allowing for electrical testing of the sensor ahead of bump-bonding. Being able to measure the sensor {$IV$ curve} before assembling a complete module can help keep the final module yields at higher levels than if this testing is restricted to fully assembled modules. Although beneficial in production, the PTB structure can also reduce charge collection efficiency of the sensor. The small depletion region around the punchthrough dots in the sensor captures some charge, resulting in a reduction of the hit efficiency in the region around the dot. In spite of this, the removal of these biasing structures is heavily disfavoured. In the measurements presented here, one device --- referred to as {DUT-A} --- has no PTB structure, whilst the other device --- referred to as {DUT-B} --- has a zigzag PTB structure variation, expected to reduce the loss of charge collection. These structures are shown in Figure~\ref{fig:PTB}.

The final production front-end chip for the ITk pixel system --- the ITkPix ---
will have $400\times384$~pixels over an area of $20.1\times21.6$~\SI{}{\mm\squared}. This will provide the resolution required throughout the ITk for precision track reconstruction. The ASIC will be a radiation hard CMOS chip with output data compression for the high radiation dose and data output requirements of the inner detector. 

For the R\&D phase, a first large-scale prototype chip, the RD53A~\cite{Garcia-Sciveres:2287593}, was produced in \SI{65}{\nm} CMOS technology by TSMC.
RD53A is the basis for production designs for the ATLAS and CMS pixel detector upgrades for the HL-LHC era.
The RD53A chip contains $400\times192$~pixels over an area of $20.1\times11.6$~\SI{}{\mm\squared}, half the area of the ITkPix. 
The RD53A chip has been designed to meet the radiation tolerance of \SI{5}{\mega\gray}, 
thinned to \SI{150}{\um}. 

Three different analogue front-ends --- \emph{linear}, \emph{differential} and \emph{synchronous} --- have been designed and implemented in the chip. The linear front-end uses a linear pulse amplification in front of the discriminator, comparing the pulse to some threshold voltage. The differential front-end uses a differential gain stage in front of the discriminator. It then implements a threshold by unbalancing the two branches. Finally, the synchronous front-end uses a baseline ``auto-zeroing'' scheme. Rather than pixel-by-pixel threshold trimming, this requires the periodic acquisition of a baseline. Detailed evaluation programmes have been carried out in both ATLAS and CMS experiments for all three front-ends to select the most suitable design for their respective operation requirements. The devices used in these tests incorporate split the chip into three regions covering equal areas on the sensor, corresponding to the three readout modes.

\section{Testbeam facilities and detector setup}
\label{sec:setup}

\begin{figure}[h!]
	\centering
	\begin{subfigure}[t]{.75\textwidth}
		\centering
		\includegraphics[width=\textwidth]{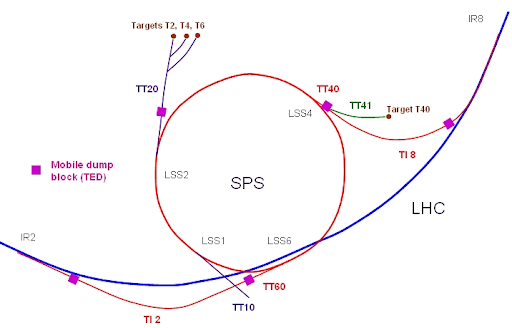}
		\caption{Schematic of the CERN SPS testbeam facility~\cite{Wenninger_2021}.}
		\label{fig:SPS}
	\end{subfigure}
	\begin{subfigure}[t]{.75\textwidth}
		\centering
		\includegraphics[width=\textwidth]{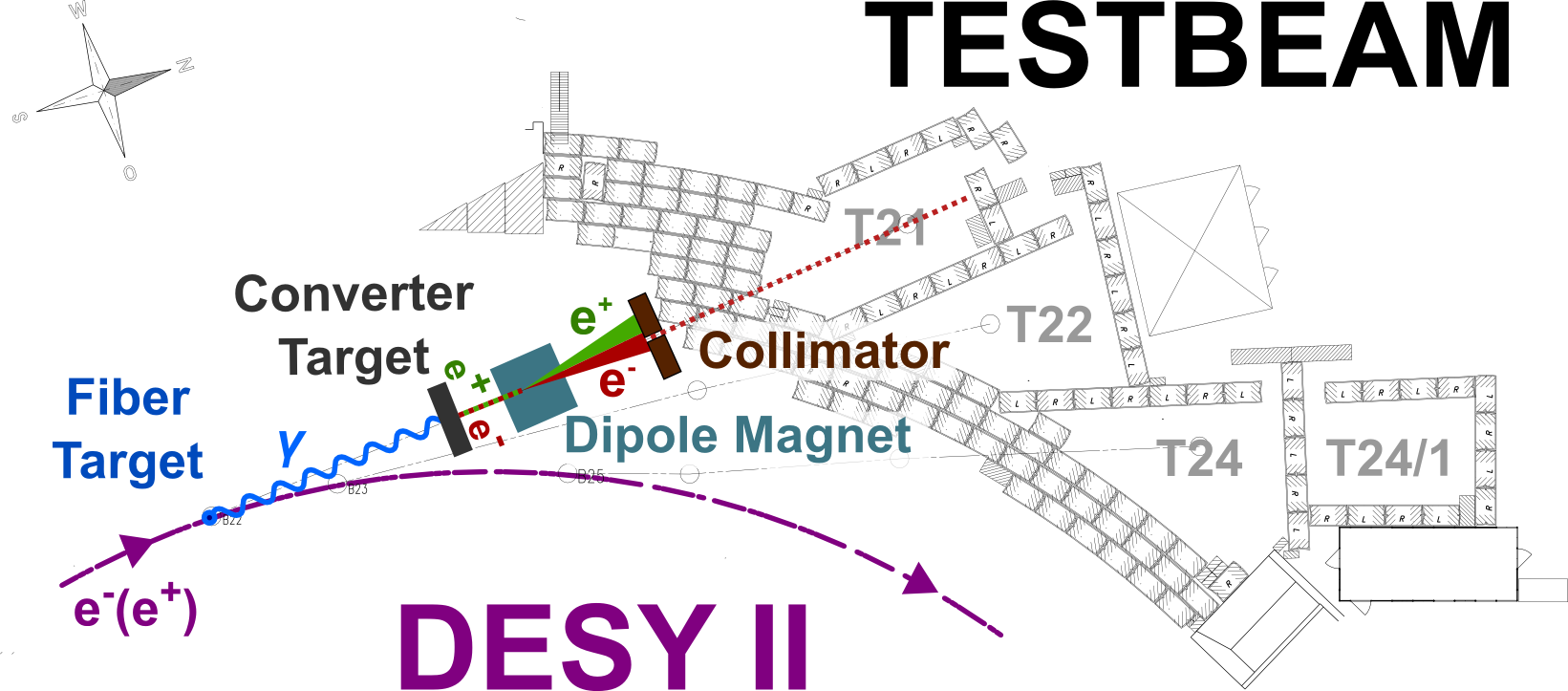}
		\caption{Schematic of the DESY testbeam facility~\cite{DESY_2021}.}
		\label{fig:DESY}
	\end{subfigure}
	\caption{
		Testbeam facilities used to make measurements of device efficiency.
	}
	\label{fig:testbeam-facilities}
\end{figure}

\paragraph{Testbeam facilities}
The testbeam campaigns considered in this paper were carried out at two facilities: the SPS testbeam facility at CERN and the DESY testbeam facility in Hamburg. The CERN SPS testbeam facility,
shown in Figure~\ref{fig:SPS}, is built around the SPS beamline~\cite{Wenninger_2021} and supplies a beam of \SI{120}{\GeV} pions from converted protons. 
The DESY testbeam facility, shown in Figure~\ref{fig:DESY}, is built around DESY-II electron-positron synchrotron and supplies a beam of \SIrange{1}{6}{\GeV} electrons from converted bremsstrahlung radiation. 
Both testbeam facilities house an EUDET-type~\cite{Jansen_2016} telescope providing identical apparatus at both sites, facilitating equivalent data reconstruction and analysis. Data was taken using the BDAQ readout system~\cite{Daas_2021}.

The difference between the two sites comes from the beam supplied. 
The nature of the beam must be considered in the analysis due to the effect of multiple scattering, which occurs due to the Coulomb forces between the atoms in the detector material and the charged particles in the beam. 
This effect is larger at lower beam-momentum and contributes to the uncertainty on the track resolution. 

\begin{figure}
\centering
\includegraphics[width=0.7\linewidth]{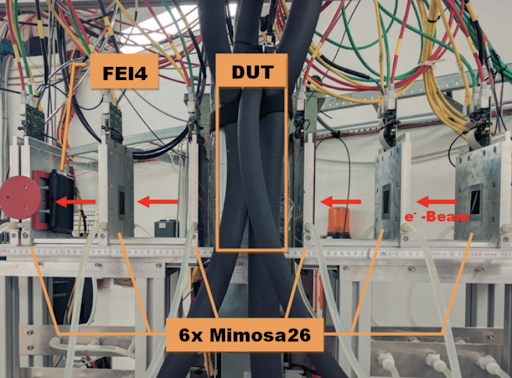}
\caption{\label{fig:TelescopeSetup}Telescope configuration at a DESY testbeam campaign. The DUT is placed between three upstream and three downstream MIMOSA-26 devices with high spatial resolution. Timing information is provided by an FE-I4 device fixed at the end of the telescope.}
\end{figure}

\paragraph{Telescope setup}
An EUDET-type beam telescope, with a setup equivalent to that shown in Figure~\ref{fig:TelescopeSetup}, is used to measure the track of a charged beam through the DUT. 

The telescope contains six MIMOSA-26~\cite{Himmi_2008} devices which are based on monolithic active pixel sensor technology with binary readout. These pixel modules have high spatial resolution which translates into a high resolution on reconstructed tracks. 
Each device covers an area of $21.5\times13.7$~\SI{}{\mm\squared} and 
has $576\times1152$~pixels with a pitch of $18.4\times18.4$~\SI{}{\um\squared} and a thickness of \SI{50}{\um}, reducing the effect of multiple scattering. 
The MIMOSA-26 devices are operated in rolling shutter readout mode with an
integration time of around \SI{115.2}{\micro\second}. This corresponds to  a readout rate of around \SI{8}{\kilo\hertz},
much slower than the \SI{40}{\mega\hertz} DUT readout rate~\cite{RD53A_manual}. This means that for a given event read from a DUT, corresponding to a single trigger, there may be several tracks reconstructed in the telescope planes, corresponding to several triggers. Removing these out-of-time contributions from the analysis is essential to measuring the DUT efficiency.
For this reason, an additional timing reference plane is placed at the end of the telescope. This timing plane is an FE-I4 pixel module~\cite{FEI4_manual}, with the same readout rate as the DUT. This FE-I4 module has pixels with pitch of $50\times250$~\SI{}{\um\squared} Reconstructed tracks to be used for the analysis of the DUT are then required to have an associated hit in the timing plane.

\paragraph{Data acquisition} 
The telescope uses dedicated trigger hardware called the trigger logic unit~(TLU). The TLU receives a signal from two scintillators placed either side of the telescope and generates triggers which it distributes to the DUT, telescope planes, and timing reference. Each trigger is uniquely timestamped, within a resolution of \SI{1.5}{\ns}~\cite{Baesso_2019}, enabling the synchronisation of hits across all planes. In addition, the DUT may respond to the TLU to indicate that it is busy, ensuring that no trigger signal is lost during the integration time. The EUDAQ software, used in conjunction with the the TLU, merges individual data streams into one and saves in histogram format for reconstruction.

\paragraph{Testbeam campaigns} 
Data was taken in October 2018 and December 2018 at the CERN SPS and DESY facilities, respectively.
Table~\ref{tab:datasets} lists the testbeam campaigns and DUTs present for each.
\begin{table}[h!]
	\caption{DUT and operating parameters for each testbeam campaign.}\label{tab:datasets}
	\begin{center}		
	\scriptsize
		\begin{tabular}{|c|c|c|c|c|c|} 
			\hline
			\emph{Campaign} & \emph{Beam}  & \emph{DUT (FE)} & \emph{PTB} &\emph{Bias voltage}~(\SI{}{\volt}) & \emph{Threshold}~(\SI{}{\electron}) \\
			\hline
			\multirow{2}{*}{CERN Oct. 2018} & \multirow{2}{*}{\SI{120}{\GeV} pions} & DUT-A (lin) & \multirow{2}{*}{none} & 600 & 1200, 1600 \\
			                                & & DUT-A (diff) & & 600 & 1160, 1680, 2140 \\
																			
			\hline
			\multirow{2}{*}{DESY Dec. 2018} & \multirow{2}{*}{\SIrange{1}{6}{\GeV} electrons} & DUT-A (lin) & none & 200, 400, 600 & 1027, 1200 \\
																			& & DUT-B (lin) & zigzag &  100, 200, 300, 400, 500, 600 & 870, 1010, 1120 \\
			\hline
		\end{tabular}
	\end{center}
\end{table}

\section{Reconstruction and analysis}\label{sec:reconstruction-analysis}
The workflow for the results presented here is summarised in Figure~\ref{fig:TB-Workflow}. The description of each stage in that workflow is given below.
\begin{figure}
	\centering
	\includegraphics[width=0.7\linewidth]{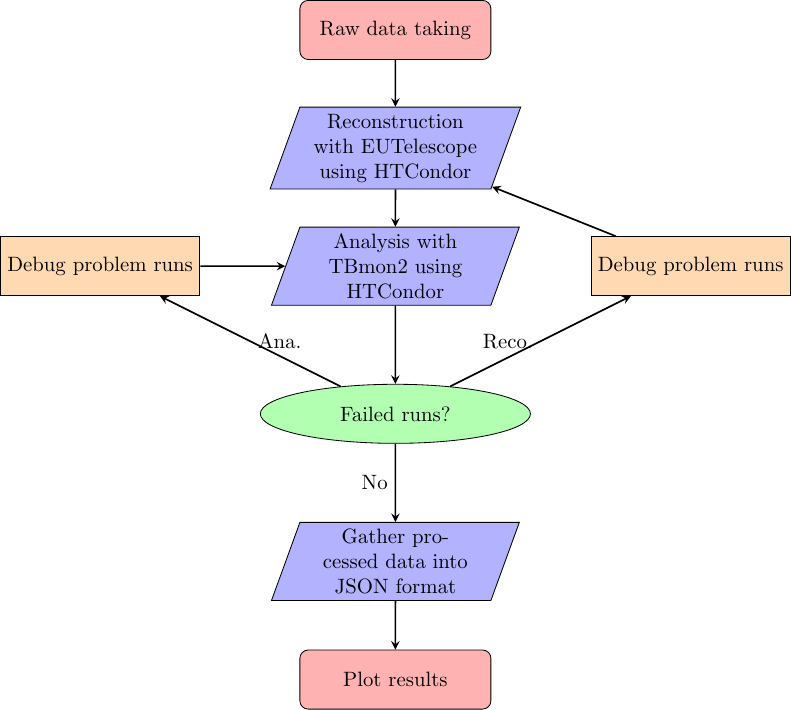}
	\caption{\label{fig:TB-Workflow}Flowchart showing the workflow for the results presented here, from the raw data-taking to final analysis of the reconstructed data.}
\end{figure}

\subsection{Reconstruction: EUTelescope}\label{sec:eutelescope}

Reconstruction of the testbeam data was carried out using EUTelescope~\cite{Bisanz_2020}, a modular framework widely used for particle trajectory reconstruction in data recorded with beam telescopes. The framework converts the raw data, as registered by the DAQ system, into a standardised data format --- LCIO~(Linear Collider Input/Output~\cite{Gaede_2003}) --- with which clusters of hit pixels can be built and assigned to particle tracks.

EUTelescope uses the GEAR~(GEometry API for Reconstruction~\cite{GEAR}) framework for the geometric description of the framework. The testbeam setup (positions, alignment, detector geometry, sensor layout, etc.) is described by an XML file which is read by the analysis framework to allow for transformations of hit positions from a local frame of reference to the telescope (global) reference frame.

\paragraph{Data conversion and noisy pixel detection}
The data recorded by the testbeam DAQ system needs to be converted from its original custom format into the standard LCIO format using EUDAQ~\cite{Ahlburg_2020}. LCIO data is event-based, containing an arbitrary amount of collections per triggered event. These collections can contain (hit) pixel indices, clustered pixels, and derived hits as the analysis moves along the reconstruction workflow.

Once the raw data has been converted into the LCIO format, one of EUTelescope processors, EUTelNoisyPixelFinder, can be applied. This uses the firing frequency~(occupancy) of the pixels to determine whether a given pixel should be labelled as noisy.
Pixels with firing frequency exceeding 0.1\% in the case of DUTs and FE-I4 reference, and 0.5\% in the case of the MIMOSA-26 planes
are identified and removed from subsequent data analysis. 

\begin{figure}[h!]
	\centering
	\begin{subfigure}[t]{.45\textwidth}
	\includegraphics[width=\linewidth]{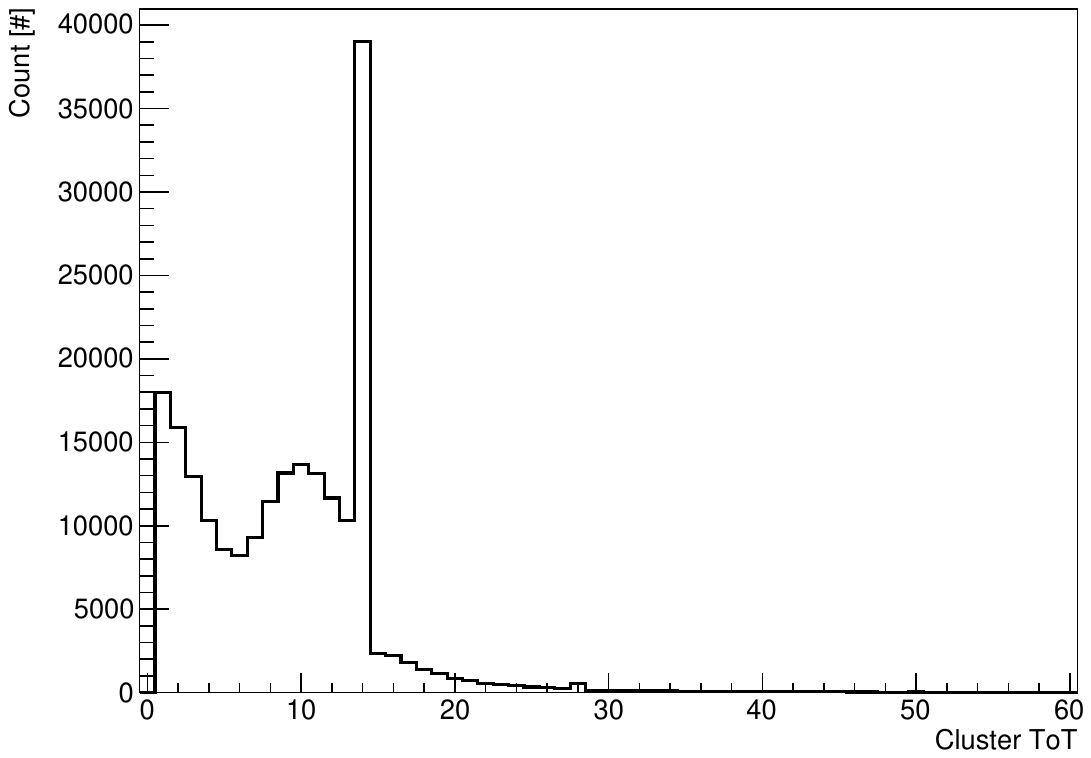}\label{fig:clustercharge}
  \caption{Cluster ToT}
  \end{subfigure}
	\begin{subfigure}[t]{.45\textwidth}
	\includegraphics[width=\linewidth]{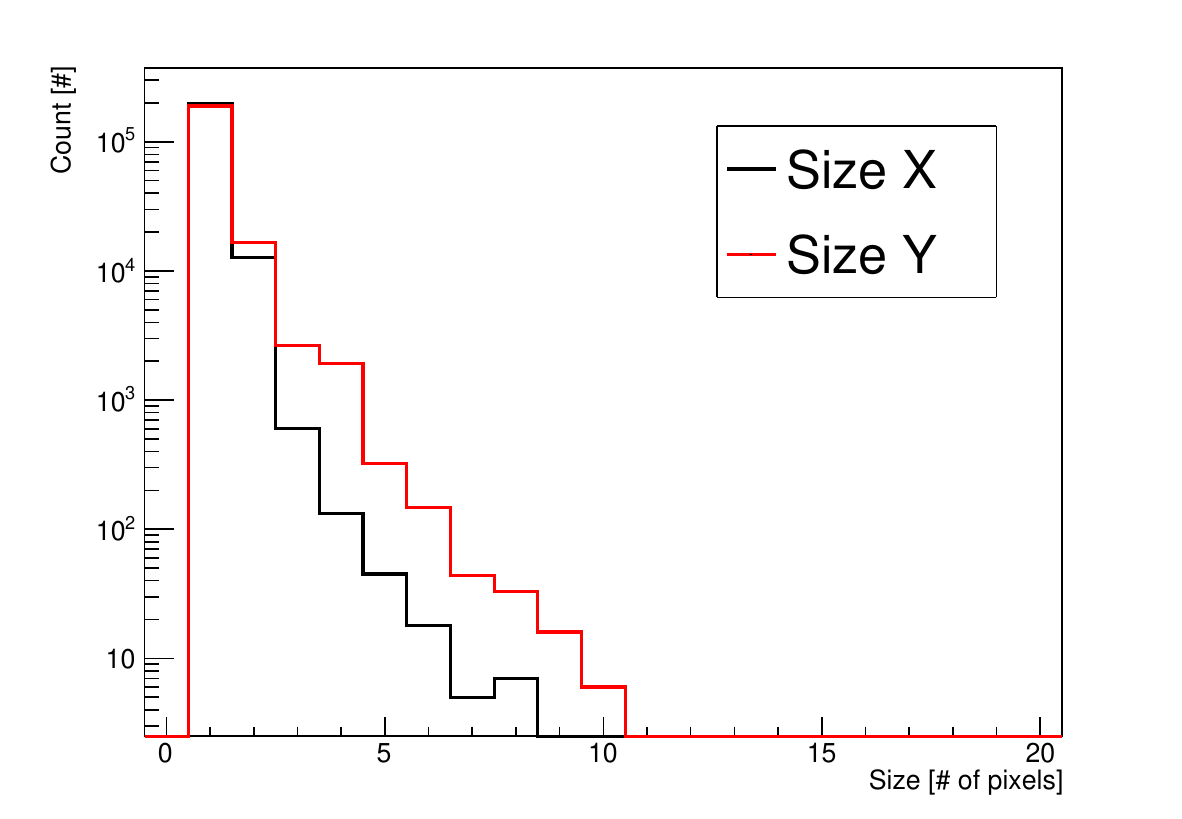}\label{fig:clustersize}
  \caption{Cluster size}
  \end{subfigure}
	\caption{
		Example run taken during DESY December 2018 testbeam with device DUT-B, $\bias=\SI{600}{\volt}$, and FE threshold of \SI{1000}{\electron}; (a) Cluster ToT in units of the beam crossing~(\SI{25}{\ns}). The exponential fall at low ToT originates from background noise. The gaussian-like peak around bin 10 represents the signal. The spike at bin 15 corresponds to noise from broken channels which send a saturated value which translates to 15~\cite{Garcia-Sciveres:2287593}. These are removed later in the reconstruction. (b) Cluster size in number of pixels. The vast majority of clusters have a size of 1 pixel.
	}
	\label{fig:cluster}
\end{figure}

\paragraph{Clustering}
Neighbouring hit pixels belonging to the same LCIO collection are grouped and stored in a new cluster collection.
Taking the masks created in the previous step, clusters containing at least one noisy pixel are tagged as noisy.
Figure~\ref{fig:cluster} shows typical distributions of total cluster charge in the form of time over threshold~(ToT).  The cluster charge is the sum of the charges of hit pixels forming the cluster. Also shown in the figure is the cluster size --- the number of hit pixels forming a cluster.

\paragraph{Hitmaker and pre-alignment}
Using the cluster collections obtained in the previous step, hit positions in the local frame of reference are defined as the centre of the
cluster coordinate, calculated as the charge-weighted centre of the position of the hits forming the cluster:
\begin{equation}
	\bar{x}=\frac{1}{Q}\sum_{i=0}^{N}x_iq_i,
	\label{eq:ClusterPosition}
\end{equation}
where $x_i$ is the position of the $i^{\text{th}}$ pixel in the cluster, $q_i$ the charge in that pixel~(ToT), and $Q$ the sum of all charges collected by all pixels in the cluster. The position is calculated in the $x$ and $y$ axes independently and stored in a new LCIO collection.

\begin{figure}[h!]
	\centering
	\begin{subfigure}[t]{.45\textwidth}
	\includegraphics[width=\linewidth]{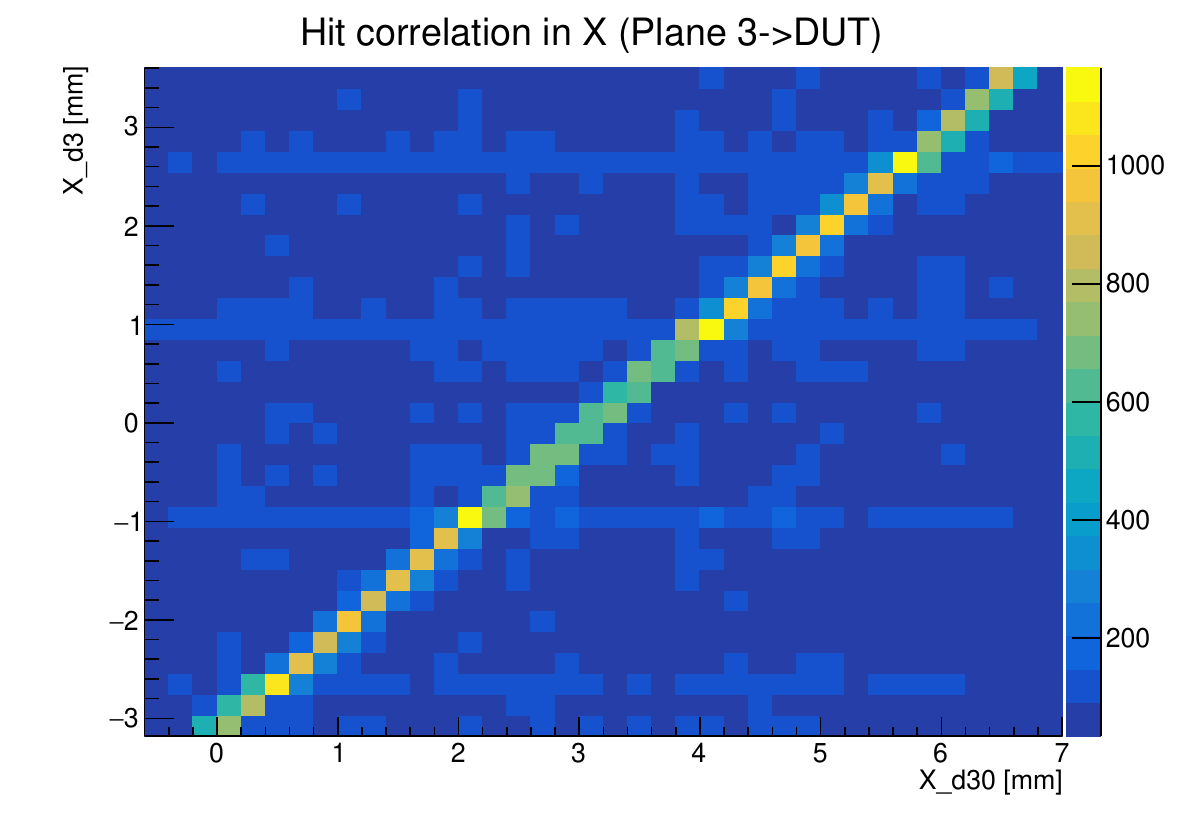}
	\caption{Correlation between hits in the $x$-direction.\label{fig:hitmaker-xcorr}}
	\end{subfigure}
	\begin{subfigure}[t]{.45\textwidth}
	\includegraphics[width=\linewidth]{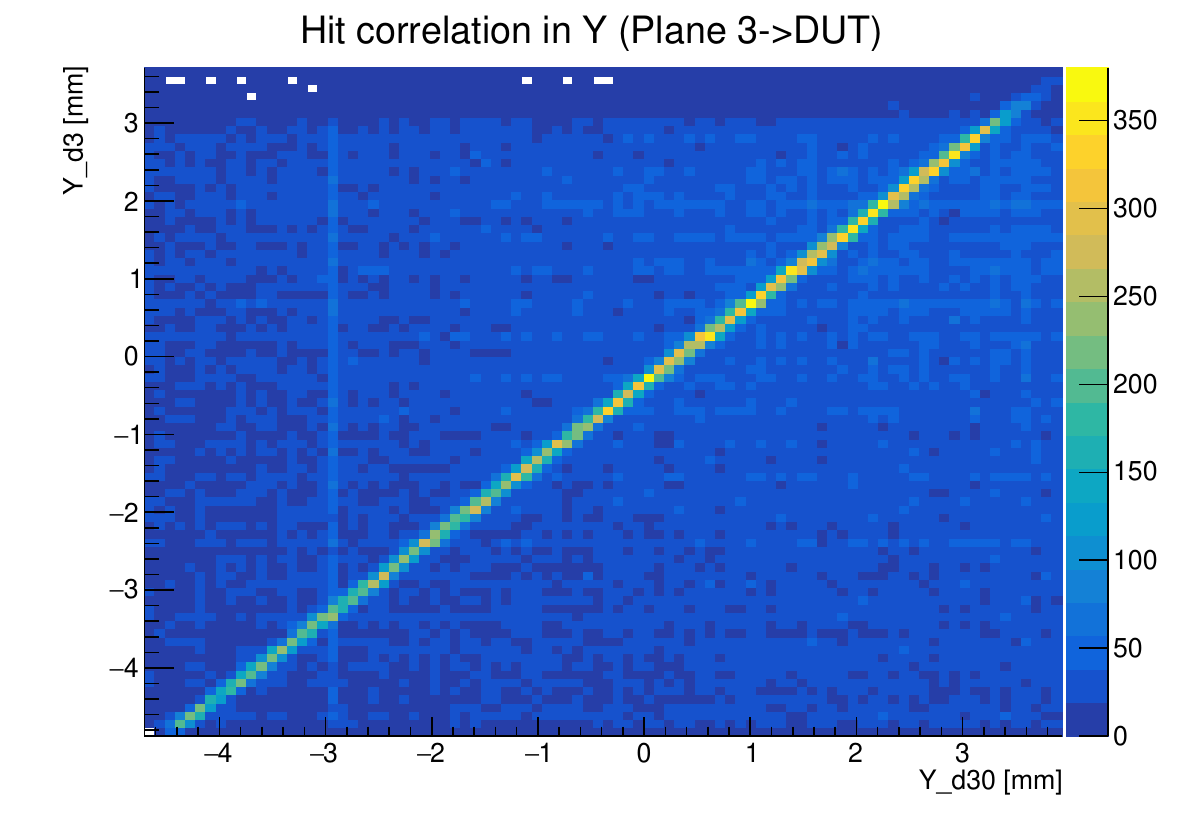}
	\caption{Correlation between hits in the $y$-direction.\label{fig:hitmaker-ycorr}}
	\end{subfigure}
	\begin{subfigure}[t]{.85\textwidth}
	\includegraphics[width=\linewidth]{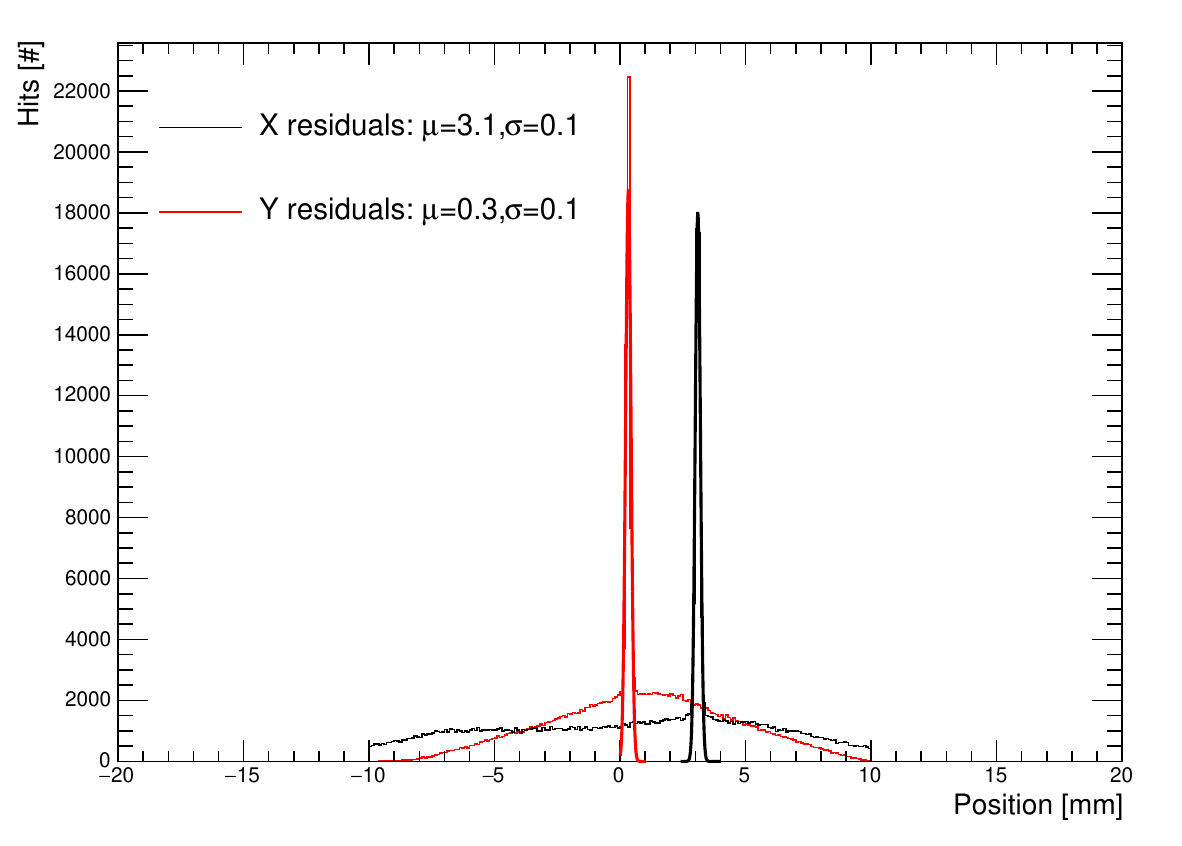}
	\caption{Residuals for both $x$ and $y$ directions.\label{fig:prealign-res}}
	\end{subfigure}
	\caption{Correlations and residuals between DUT and adjacent telescope plane. The device used in this example has pixel pitch $25\times100$~\SI{}{\um\squared}, visible in the difference in granularity in the two dimensions. Those used in the measurements presented in the following sections all have pitch of $50\times50$~\SI{}{\um\squared}.}
	\label{fig:Correlations}
\end{figure}

The hit positions are translated from the local frame of reference for each plane to a global frame by rotating and shifting the local coordinates by the angles and global positions registered when performing data-taking (i.e. rotation and position of each plane and DUT in the testbeam).
These values are provided by the GEAR file. Once the hits are provided in a global frame of reference, the positions of those hits registered in the first sensor are propagated to all sensors, calculating the difference between the propagated value and the registered hit in each sensor,
referred to as residuals.
These residuals are registered in one and two-dimensional histograms,
taking the bins with highest counts as the pre-alignment factors, representing
a rough estimate of the shift of the planes in $x$ and $y$ direction.
These pre-alignment values are then written to a new GEAR file to be used in subsequent steps.
Examples of correlations of the hit positions in $x$ and $y$ between
hits in the DUT and a beam telescope plane are shown in Figure~\ref{fig:Correlations}, where (c) shows the single dimension projection of (a) and (b) overlaid.
The distance of the peak of each projection from the origin represents the displacement of the DUT from its optimum position relative to the telescope planes.
This value is used to align plains.

\paragraph{Alignment}
In this step, updated global hits based on the corrected pre-aligned GEAR file
are used to determine precise alignment of the telescope and DUT planes,
using track finding and aligning algorithms based on
General Broken Lines (GBL) track model~\cite{Kleinwort_2012}.
In this model, the beam telescope, consisting of six planes, is divided into two groups -- the upstream and downstream triplets.
Within each triplet, a straight line~(doublet) is calculated joining the first and last hits.
Doublets that have slopes inconsistent with the beam direction are excluded to suppress false combinations.
The threshold is given by user-defined cuts.
The distance between the doublet and the hit in the middle sensor must also be within a user-defined range for the triplet to be considered valid.
Both upstream and downstream triplets are then extrapolated to
the centre of the beam telescope.
The extrapolated position of these two triplets must be matched within
a user-defined distance to be joined together as a track.
Once the fitted line from each arm of telescope is matched the track is defined.

The alignment process is repeated 3 times per run, using a partial number of events to retrieve and estimate the values of the cuts to be used by GBL.
As the alignment process is iterated, cut values are reduced, initially set high for a first estimate of the cut values,
then optimised by performing a gaussian fit on the distribution of the distances and slopes previously explained. 

\begin{figure}[h!]
	\centering
	\begin{subfigure}[t]{.45\textwidth}
	\includegraphics[width=\linewidth]{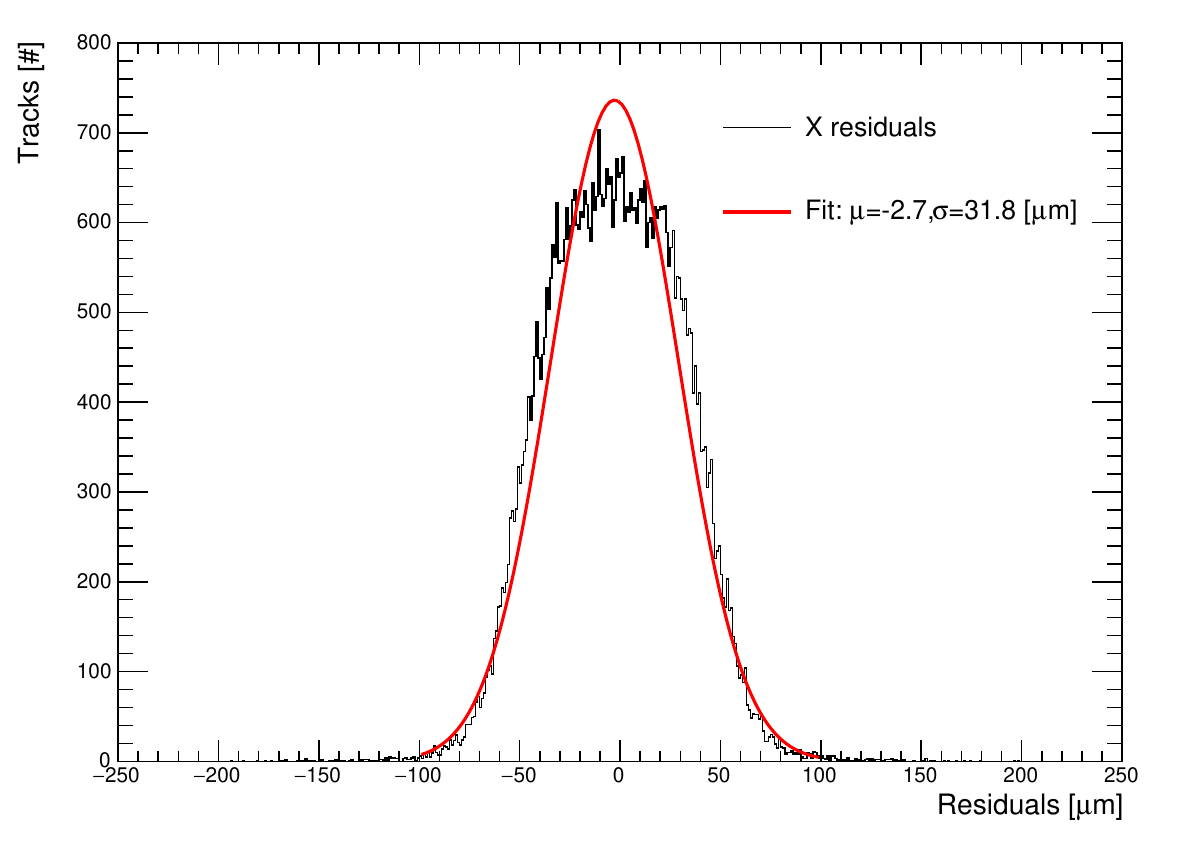}
	\end{subfigure}
	\begin{subfigure}[t]{.45\textwidth}
	\includegraphics[width=\linewidth]{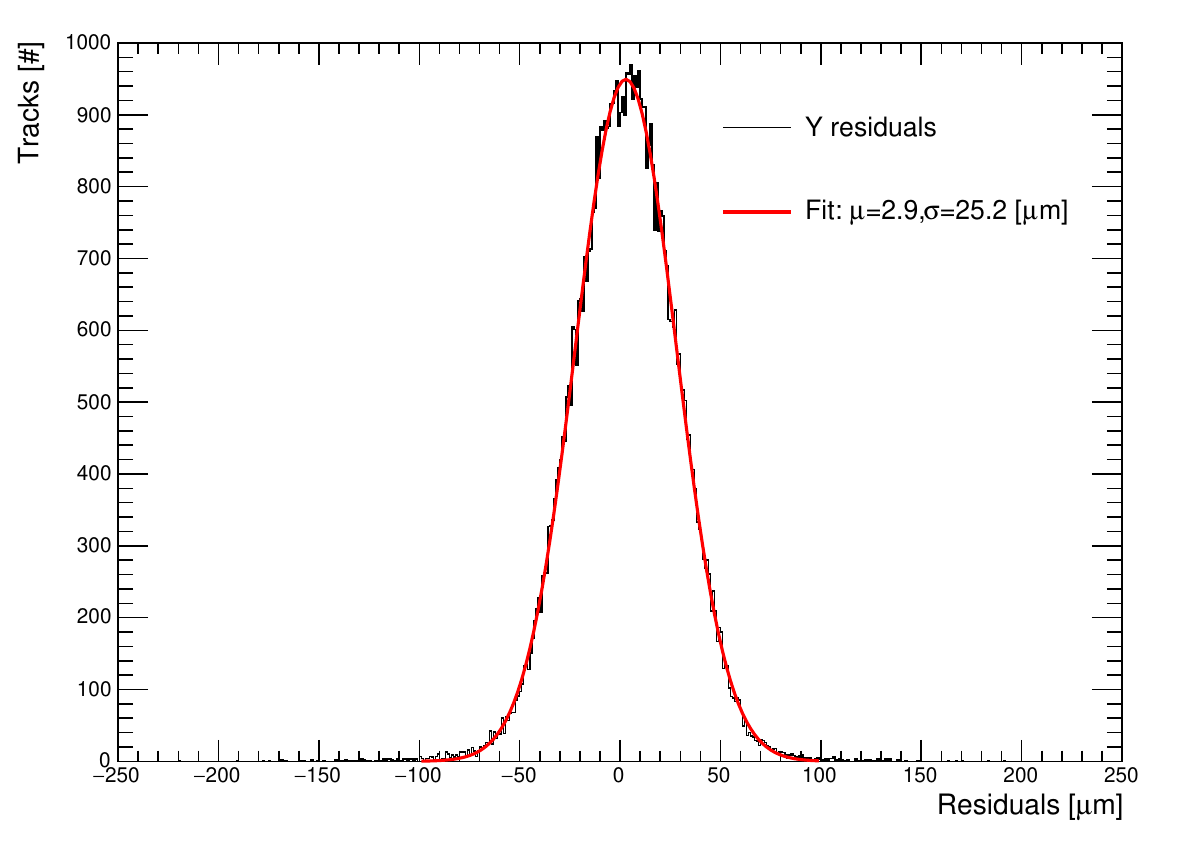}
	\end{subfigure}
	\caption{
	Residual distributions after the track fitting.}
	\label{fig:Residuals}
\end{figure}

\paragraph{Track fitting}
The final reconstruction step consists of a track fit using the six hits of the telescope planes associated to each track found in the previous step passing all required cuts. The tracks are output to a ROOT $n$-tuple so that they may be used for subsequent analysis.
The resulting residuals for the hits in DUT in $x$ and $y$,
defined as the difference between the reconstructed
cluster position and the extrapolated values from the fitted track are shown in Figure~\ref{fig:Residuals}.
Both distributions are centred around zero, which shows good alignment with no systematic offset.
The RMS of each distribution corresponds to the expected resolutions of the pixel pitches in each dimension.
The $\chi^{2}$ per degree of freedom distribution for the track fit peaks at low values and has a smooth tail, implying that the telescope is well aligned over the run.

\subsection{Analysis: TBmon2}\label{sec:tbmon2}

The output files produced following reconstruction in
EUTelescope are then analysed using TBmon2~\cite{tbmon2}, a testbeam analysis software package. 
The framework uses a core processor to read all relevant inputs,
such as data paths and DUT geometry files,
and then executes data pre-processing and different analyses. 
These analyses can be individually configured in dedicated configuration files with
corresponding DUT specifications.
Central configuration parameters, such as selection of
fiducial regions and track quality criteria, are set by the user.

The pre-processing module first finds tracks that pass a set of basic quality criteria and
are matched to at least one hit in the timing reference plane.
When two DUTs with the same readout rate are placed within the telescope planes,
it is possible to use either the other RD53A DUT or the FE-I4 as the timing reference plane.
To mitigate multiple scattering effect when propagating tracks from
the DUT plane to the reference plane, a radius matching requirement
is applied on hits from both devices.
The maximum distance between hits in each axis in the transverse plane is specified in the configuration
file and is commonly set to half of the pitch of the DUT in that direction.

Once the pre-processing is complete, different analysis modules can be executed to study the performance of the DUT.
In this paper, we focus on the pixel hit efficiency, defined as the ratio of the number of tracks with matching
hits in the DUT to the total number of telescope tracks:
\begin{equation}
	\epsilon = \frac{n_{\text{matched tracks}}}{n_{\text{total tracks}}}.
	\label{eq:efficiency}
\end{equation}
This quantity is one of the most important figure of merit to qualify sensor designs.
During the ATLAS ITk Pixel Sensor Market Survey~\cite{M_bius_2022}, devices achieving global efficiencies above 
98.5\% or 97\% in the case of unirradiated or irradiated modules, respectively,
were considered to have met the production requirements.

All tracks passing the quality selection criteria applied in the pre-processing are included
in the total number of tracks.
Hit-to-track matching criteria is specified by maximum matching distance in the
$x-y$ plane between the extrapolated position in the DUT plane from the telescope tracks and
the reconstructed pixel hit position in the DUT.
The maximum threshold in each axis, $x$ or $y$, is set to be twice
the pitch size in the given axis of the DUT and specified in the configuration file.

\begin{figure}[h!]
	\centering
	\begin{subfigure}[t]{.43\textwidth}
	\includegraphics[width=\linewidth]{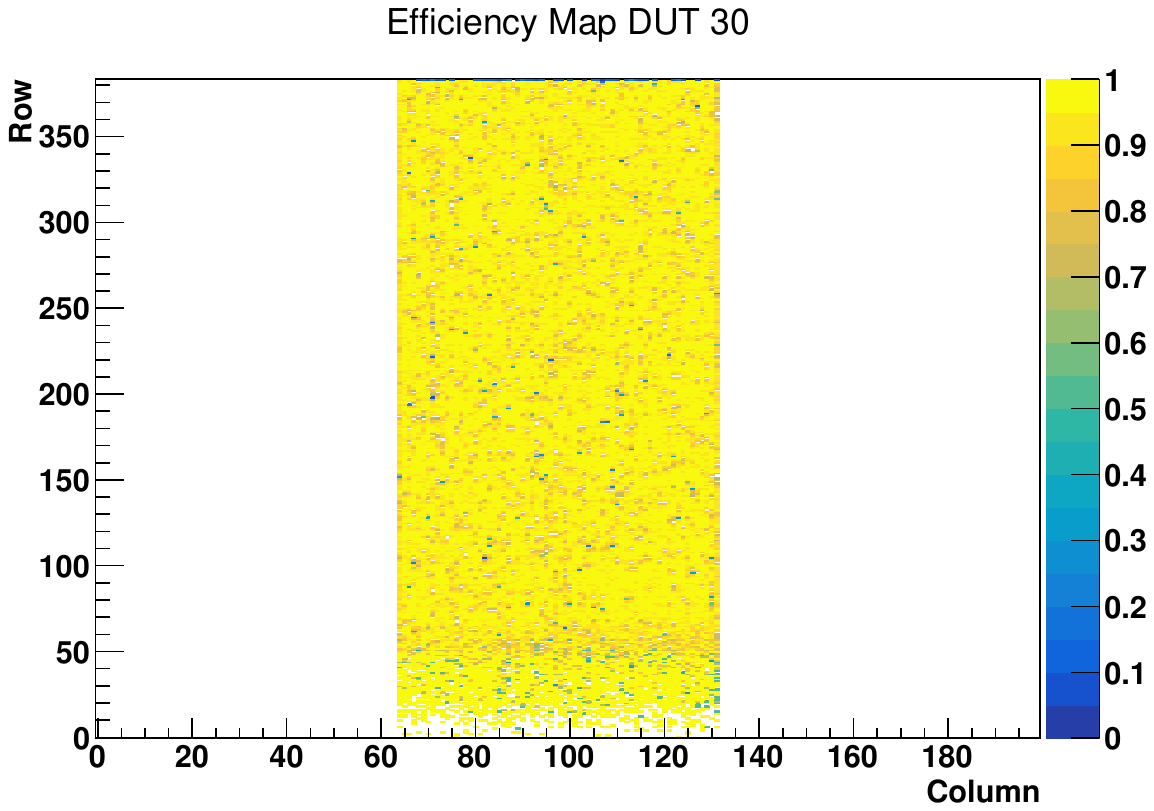}\label{effmap_2D}
	\caption{Typical pixel efficiency map using one of three front-ends on the RD53A.}
	\label{fig:effmap_2D}
	\end{subfigure}
	\begin{subfigure}[t]{.51\textwidth}
	\includegraphics[width=\linewidth]{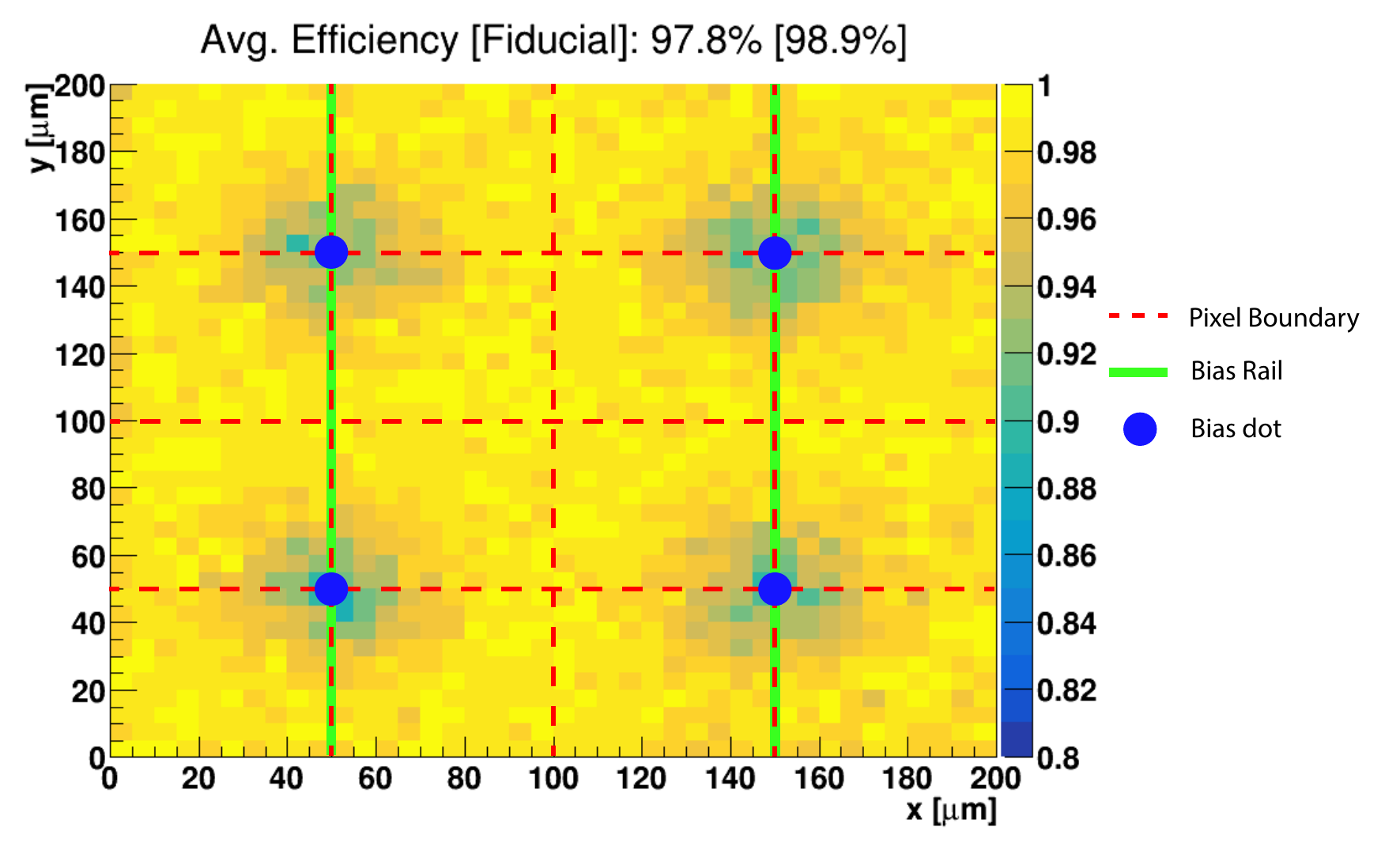}
	\caption{Typical in-pixel efficiency map for a device including punch-through bias.}
	\label{fig:inpix_effmap_2D}
	\end{subfigure}
	\caption{Example run during the December 2018 testbeam campaign (Batch 1, DUT-B, {$\bias=\SI{600}{\volt}$}, {$\text{threshold}=\SI{1000}{\electron}$)}.
	}
	\label{fig:EffMap}
\end{figure}

In addition to the overall pixel efficiency, more detailed studies are also
carried out to characterise the pixel hit efficiency dependence on position within the
pixel matrix:
\begin{itemize}
	\item Pixel hit maps: the efficiency is computed on a pixel-by-pixel basis by considering all hits and tracks which pass through that pixel. The outputs of this analysis are efficiency maps such as that shown in Figure~\ref{fig:effmap_2D}, where the efficiency is only computed for the front-end under study (columns 65-130).

	\item In-pixel efficiency: this analysis provides an in-depth look at the efficiency patterns that may arise due to different structures present in the device at a sub-pixel level, such as punch-through bias dots. The ability to make such a measurement is only possible as a result of the high resolution of the MIMOSA-26 sensors in the telescope in comparison to the DUTs. In order to analyse the efficiency in such a granular manner, a large number of hits per pixel are required. To achieve this, and avoid the need for extremely long data-taking runs, the full pixel matrix is divided into blocks containing $4\times4$~pixels each, leading to a total block area of $200\times200$~\SI{}{\um\squared}, with each block across the whole pixel matrix then overlaid on top of one another  to provide averaged information from across the whole sensor.
	The resulting efficiency maps are shown in Figure~\ref{fig:inpix_effmap_2D}, where it is possible to distinguish efficiency drops between pixels corresponding to the presence of a punch-through bias dot, which is at the same potential as the charge collection electrode and tends to collect the electrons around it rather than the pixels themselves.
	A fiducial area can be calculated by masking pixels in the region affected by the punch-through dots, providing a fiducial efficiency as well as the overall efficiency. With reference to Figure~\ref{fig:inpix_effmap_2D}, this region is defined in Equation~\ref{eq:fiducial}:
	\begin{equation}
	(x, y) \in \left( [0\, \SI{}{\um}, 20\, \SI{}{\um}) \cup (80\, \SI{}{\um}, 120\, \SI{}{\um}) \cup (180\, \SI{}{\um}, 200\, \SI{}{\um}] \right)^2.
	\label{eq:fiducial}
	\end{equation}
\end{itemize}

\clearpage

\section{Results}\label{sec:results}
Presented in the following sections are results of the final analysis for both testbeam campaigns, as well as a combined set of results comparing devices across campaigns. The criteria later adopted by the ATLAS planar sensor market survey required that devices irradiated to \SI{2e15}{\nequ\per\cm\squared} achieve 97\% efficiency at a bias voltage of $\bias=\SI{400}{\volt}$ and that devices irradiated to \SI{5e15}{\nequ\per\cm\squared} achieve 97\% efficiency at a bias voltage of $\bias=\SI{600}{\volt}$. Although these criteria were adopted after the irradiation of the devices measured here, they provide a useful reference point for the expected performance post-irradiation. Where possible, devices are evaluate with reference to the former benchmark and are therefore conservative in that the results presented here are for devices irradiated to \SI{3.4e15}{\nequ\per\cm\squared} --- almost double the fluence stipulated for that set of operating parameters. Efficiency values are also evaluated for various other sets of parameters. The statistical uncertainties on the results shown here are too small to be visible on the plots, so are omitted.

\subsection{October 2018 testbeam}
\label{sec:results_oct2018}

In this campaign, device DUT-A was characterised using both linear and differential front-ends. It contains no punch-through biasing structure.

\begin{figure}[!htb]
	\centering
	\includegraphics[width=\linewidth]{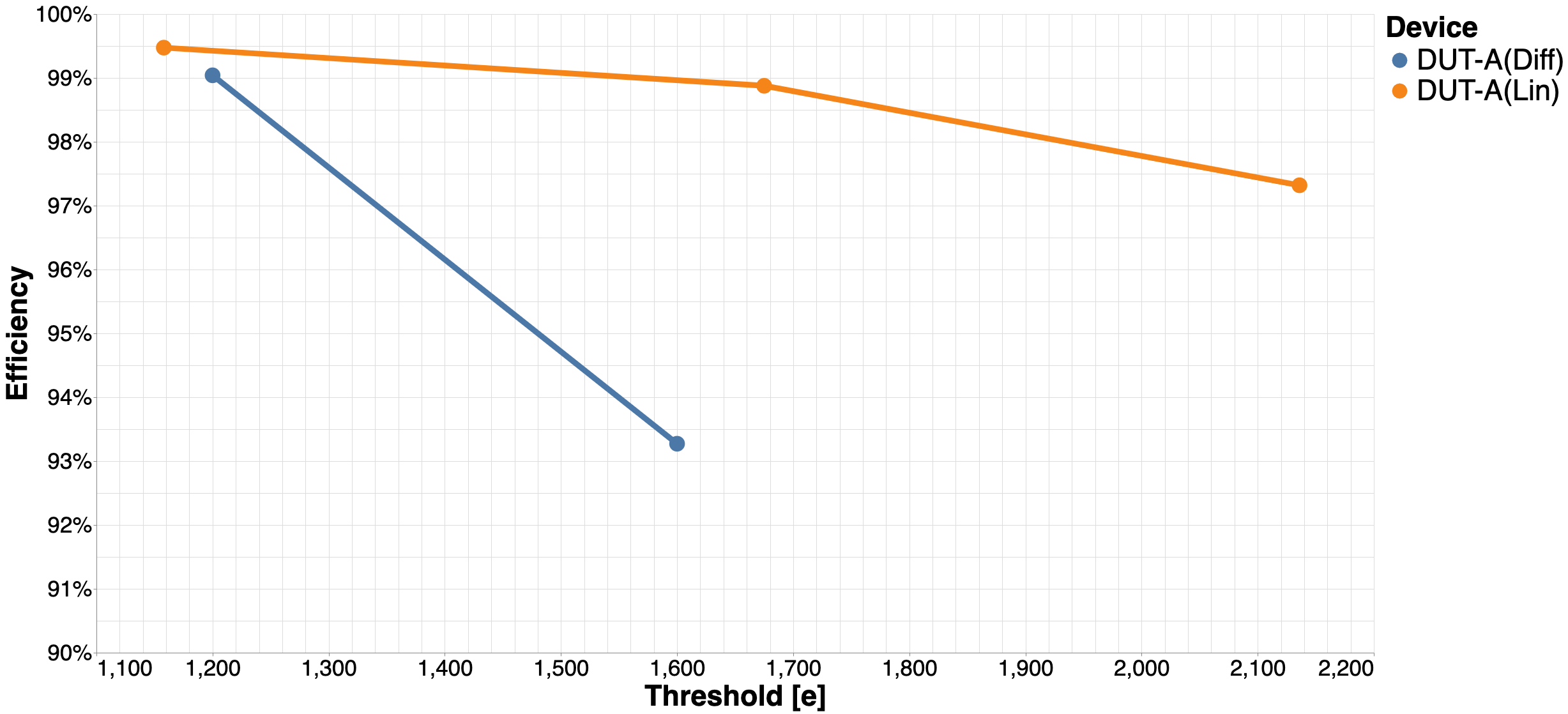}
	\caption{Evolution of the efficiency~(as defined in Section~\ref{sec:eutelescope}), as a function of threshold, in the October 2018 testbeam for both front-ends at $\bias=\SI{600}{\volt}$.}
	\label{fig:results:Oct18 eff vs thl}
\end{figure}

Figure~\ref{fig:results:Oct18 eff vs thl} shows the overall efficiency as a function of the threshold for the
differential and linear FEs, at a $\bias=\SI{600}{\volt}$.
The highest efficiency of $\varepsilon=99.47\%$ is observed with a threshold of  \SI{1158}{\electron} for the linear FE, 
and $\varepsilon=99.05\%$ for the differential FE at a threshold of \SI{1200}{\electron}. 
Increasing the threshold beyond this point, the devices using both of the FEs exhibit a degradation in efficiency. 
This can be attributed to the loss of signal.

\begin{figure}[!htb]
	\centering
	\begin{subfigure}[t]{.48\textwidth}
		\includegraphics[width=\linewidth]{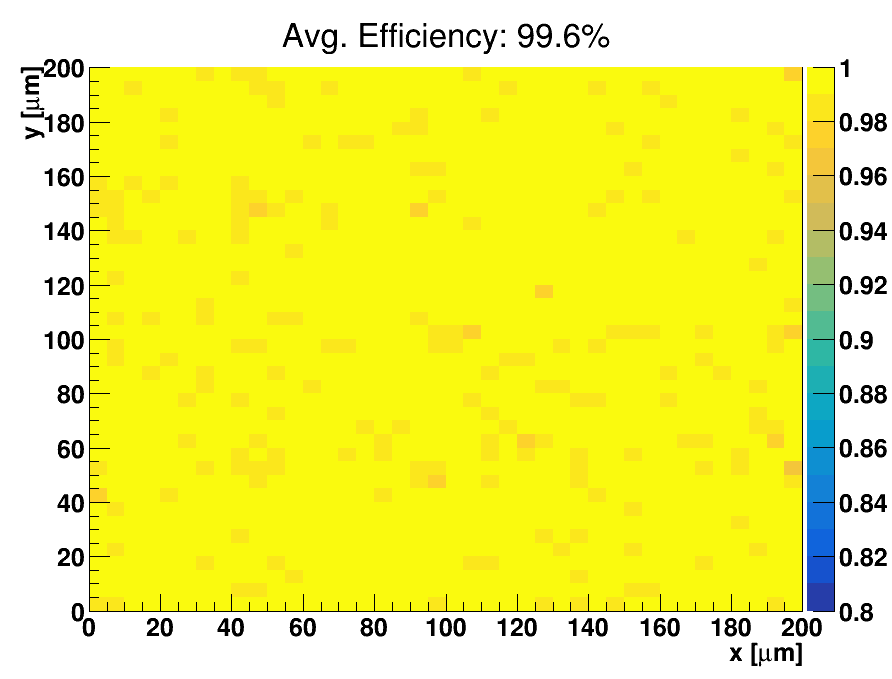}
		\caption{$\text{threshold}=\SI{1158}{\electron}$}
	\end{subfigure} \hfill%
	\begin{subfigure}[t]{.48\textwidth}
		\includegraphics[width=\linewidth]{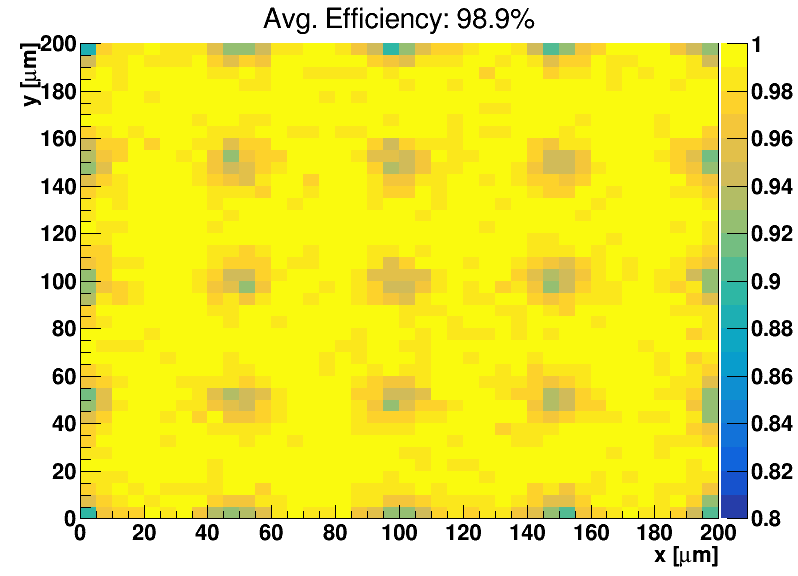}
		\caption{$\text{threshold}=\SI{1675}{\electron}$}
	\end{subfigure}
	\begin{subfigure}[t]{.48\textwidth}
		\includegraphics[width=\linewidth]{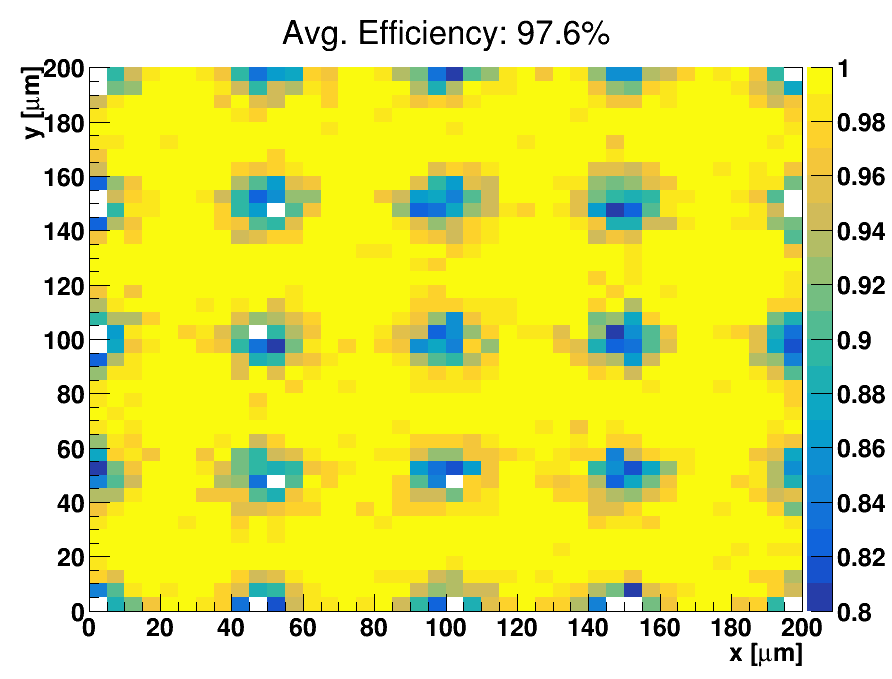}
		\caption{$\text{threshold}=\SI{2136}{\electron}$}
	\end{subfigure} \hfill%
	\caption{In-pixel efficiency maps, as defined in Section~\ref{sec:tbmon2}, for device DUT-A (linear FE) with $\bias=\SI{600}{\volt}$.}
	\label{fig:results:Oct18 in-pixel eff Lin V600}
\end{figure}

\begin{figure}[!htb]
	\centering
	\begin{subfigure}[t]{.48\textwidth}
		\includegraphics[width=\linewidth]{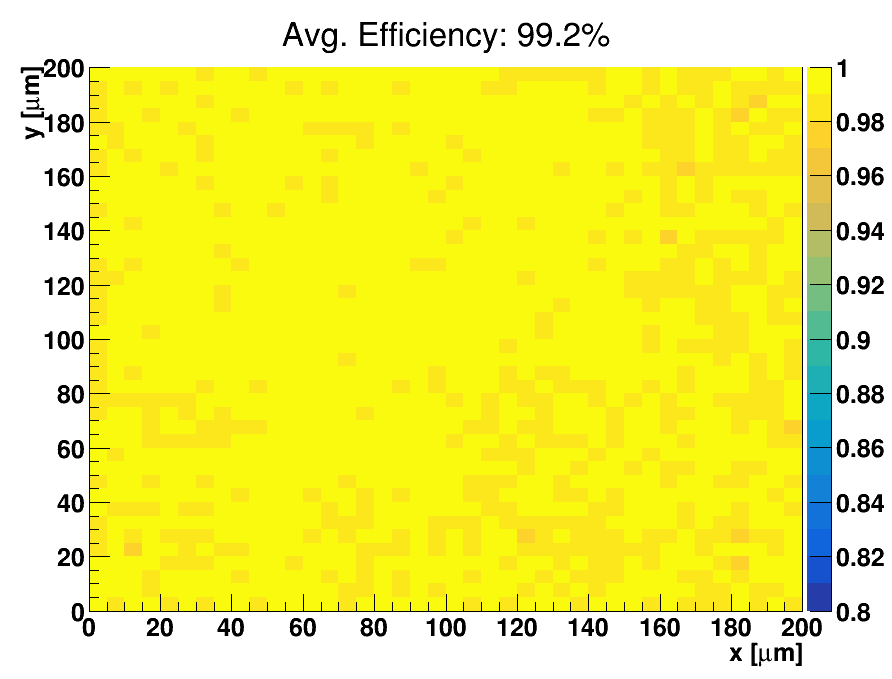}
		\caption{$\text{threshold}=\SI{1200}{\electron}$}
	\end{subfigure} \hfill%
	\begin{subfigure}[t]{.48\textwidth}
		\includegraphics[width=\linewidth]{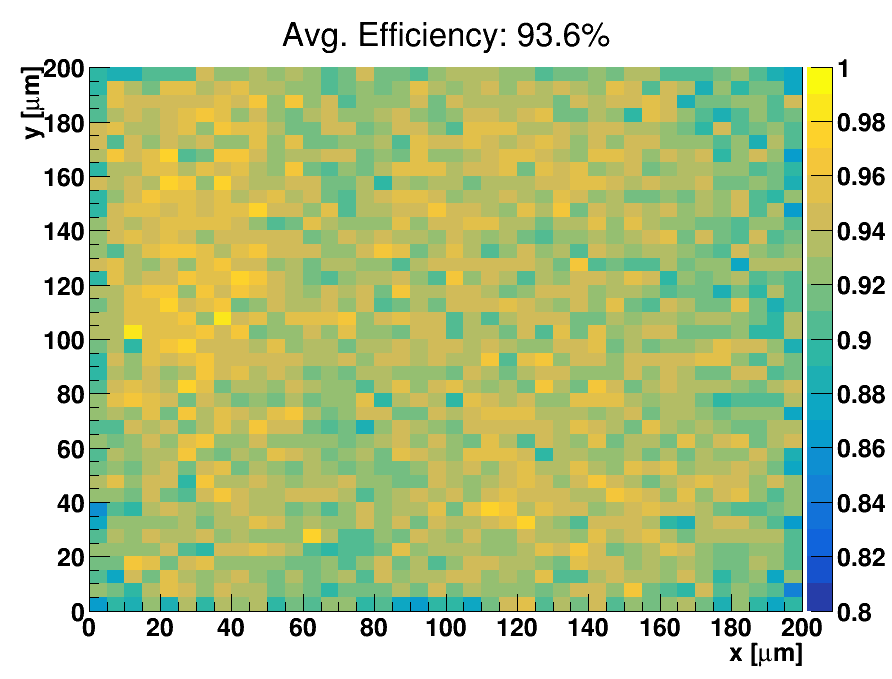}
		\caption{$\text{threshold}=\SI{1600}{\electron}$}
	\end{subfigure}
	\caption{In-pixel efficiency maps for device DUT-A (differential FE) with $\bias=\SI{600}{\volt}$.}
	\label{fig:results:Oct18 in-pixel eff Diff V600}
\end{figure}

Figure~\ref{fig:results:Oct18 in-pixel eff Lin V600} shows the in-pixel efficiency maps
for the linear FEs at different thresholds. At the optimum threshold, the efficiency exhibits near-uniform distribution across the sensor. At higher threshold settings, efficiencies are reduced close to the pixel corner boundaries as a result of charge-sharing. 
Figure~\ref{fig:results:Oct18 in-pixel eff Diff V600} shows the in-pixel efficiency maps
for the differential FEs at different thresholds. It can be seen that the device efficiency drops as the threshold is raised beyond the efficiency plateau such that signal is being cut away.

\clearpage

\subsection{December 2018 testbeam}
\label{sec:results_dec2018}

In this campaign, both devices DUT-A and DUT-B were characterised using the linear front-end. DUT-A contains no PTB structure, whilst DUT-B uses the zigzag PTB structure.

\begin{figure}[!htb]
	\centering
	\includegraphics[width=\linewidth]{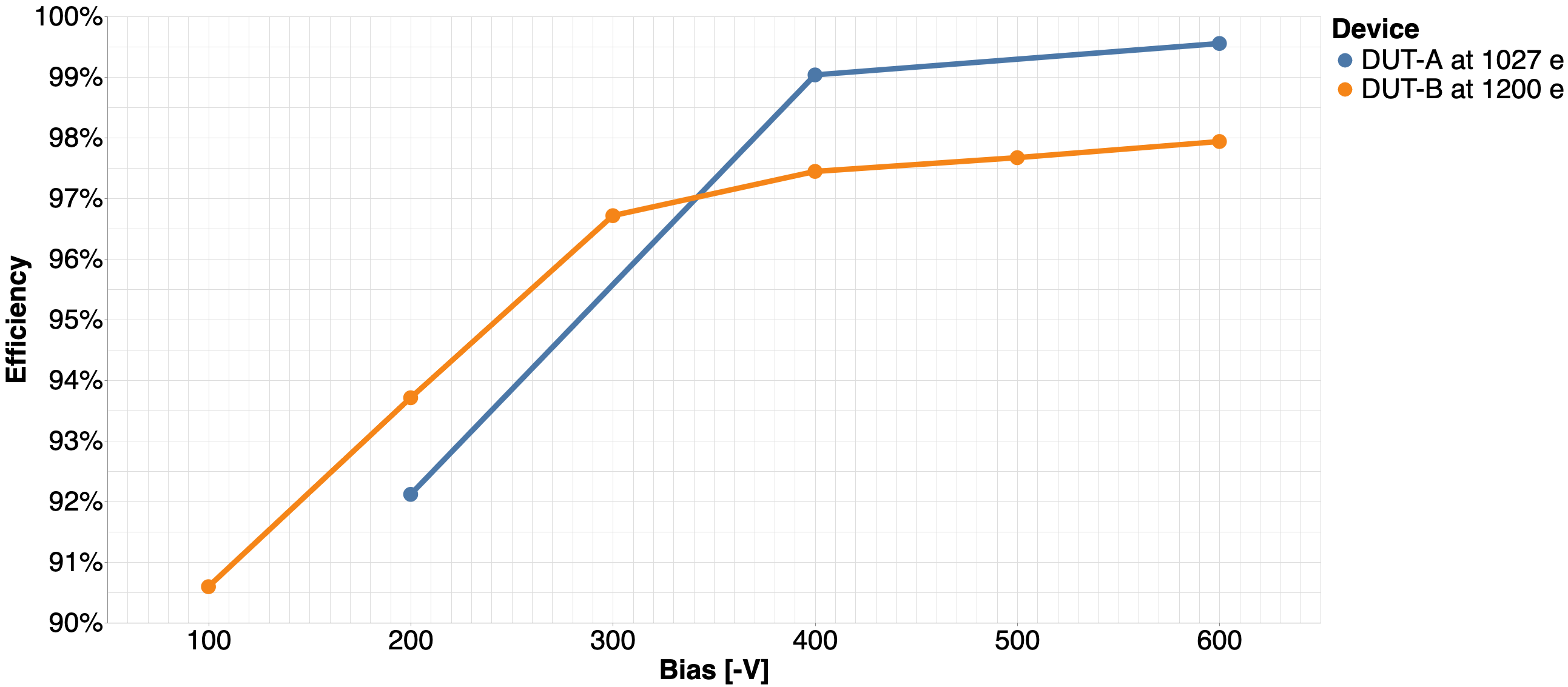}
	\caption{Evolution of the efficiency as a function of bias voltage during the December 2018 testbeam with $\text{threshold}=\SI{1200}{\electron}$ (DUT-B) and $\text{threshold}=\SI{1027}{\electron}$ (DUT-A). Both devices are using the linear front-end.}
	\label{fig:results:Dec18 eff vs bias}
\end{figure}

\begin{figure}[!htb]
	\centering
	\includegraphics[width=\linewidth]{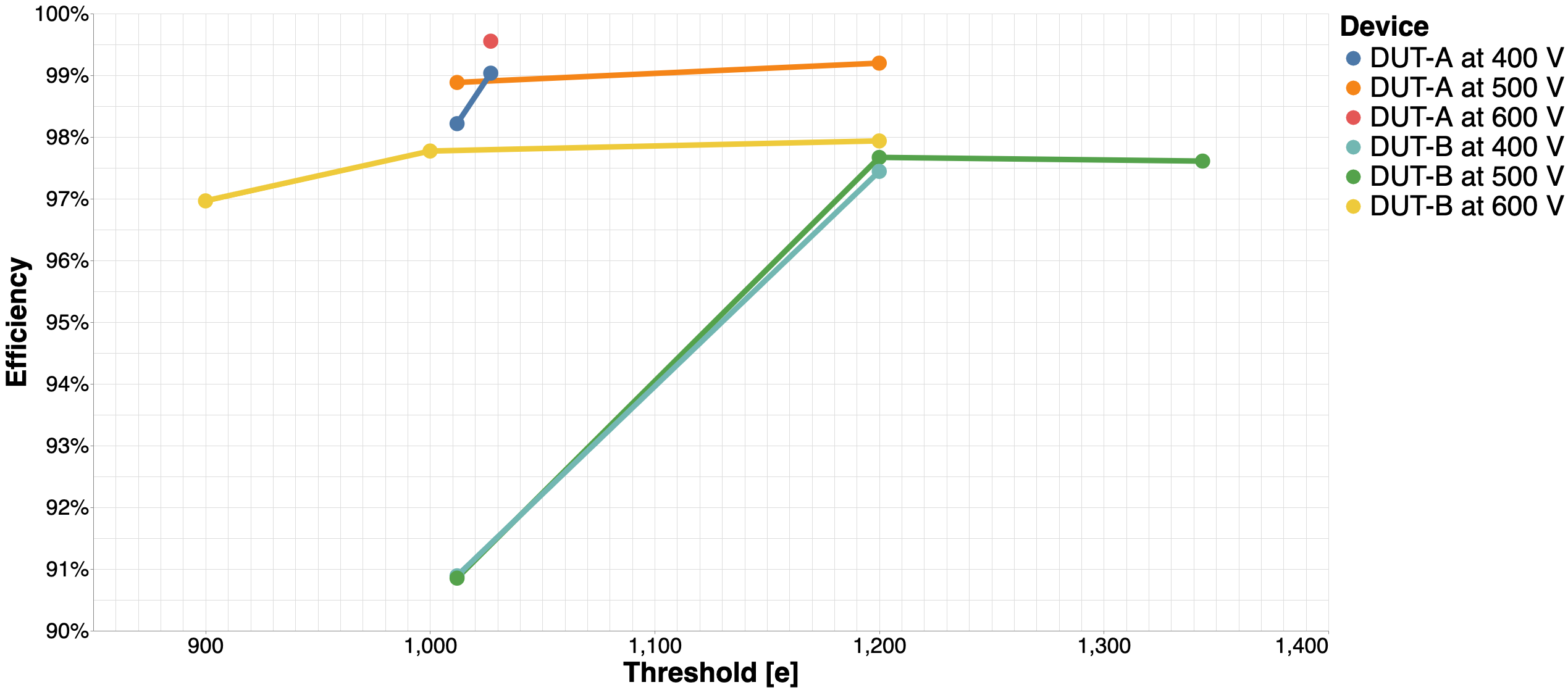}
	\caption{Evolution of the efficiency as a function of threshold during the December 2018 testbeam for both devices, using the linear front-end.}
	\label{fig:results:Dec18 eff vs thl}
\end{figure}

Figure~\ref{fig:results:Dec18 eff vs bias} shows the overall efficiency as a function
of the $\bias$ for both DUTs. In such comparisons, the efficiency values displayed for DUT-B are the global efficiencies, as opposed to those for the fiducial area previously defined in Equation~\ref{eq:fiducial}.
Both devices behave as expected, with efficiency increasing as $\bias$ increases before reaching a plateau.
The efficiency reaches plateau by $\bias=\SI{300}{\volt}$ and $\bias=\SI{400}{\volt}$ for
device DUT-B and DUT-A, respectively.
Figure~\ref{fig:results:Dec18 eff vs thl} shows the overall efficiency as a function
of the threshold for both DUTs.
The highest efficiency achieved for device DUT-B is observed with $\text{threshold}=\SI{1013}{\electron}$ at $\bias=\SI{600}{\volt}$. An efficiency of $\varepsilon=99.55\%$ is reached for DUT-A at $\text{threshold}=\SI{1027}{\electron}$ at $\bias=\SI{600}{\volt}$.
At $\bias=\SI{400}{\volt}$, both devices exceed the 97\% efficiency criterion.
Device DUT-B reaches the desired efficiency when using a threshold above $\SI{1000}{\electron}$.
For device DUT-A, the desired efficiency is reached for all threshold points with $\bias\geq\SI{400}{\volt}$.


\begin{figure}[!htb]
	\centering
	\begin{subfigure}[t]{.48\textwidth}
		\includegraphics[width=\linewidth]{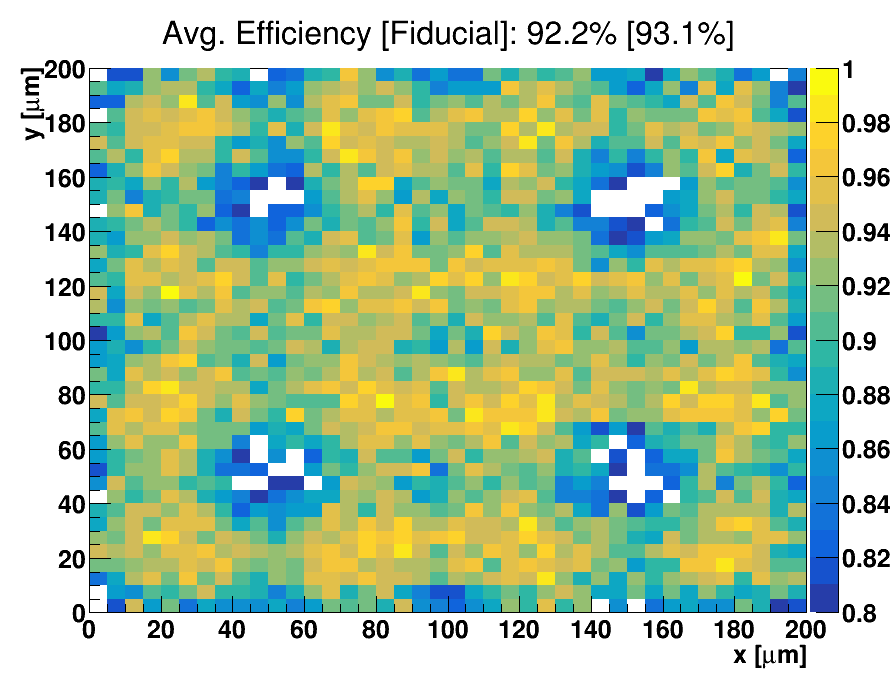}
		\caption{$\bias=\SI{100}{\volt}$}
	\end{subfigure} \hfill%
	\begin{subfigure}[t]{.48\textwidth}
		\includegraphics[width=\linewidth]{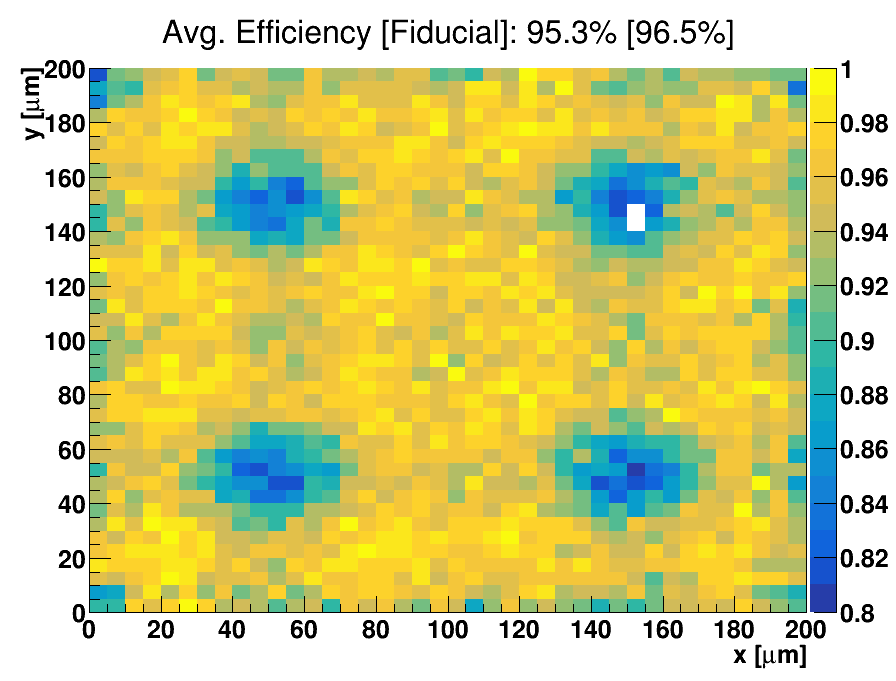}
		\caption{$\bias=\SI{200}{\volt}$}
	\end{subfigure}
	\begin{subfigure}[t]{.48\textwidth}
		\includegraphics[width=\linewidth]{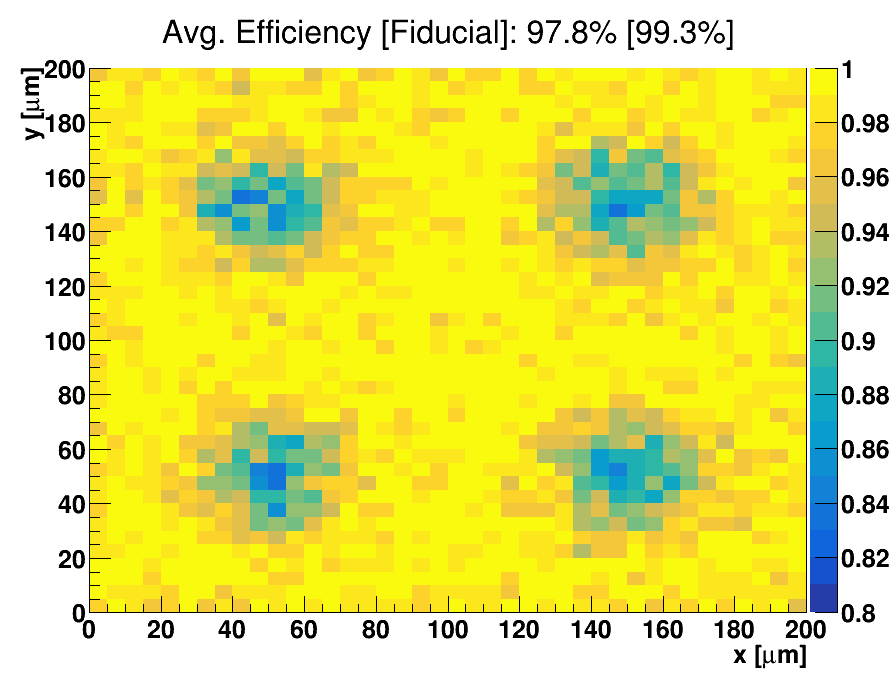}
		\caption{$\bias=\SI{300}{\volt}$}
	\end{subfigure} \hfill%
	\begin{subfigure}[t]{.48\textwidth}
		\includegraphics[width=\linewidth]{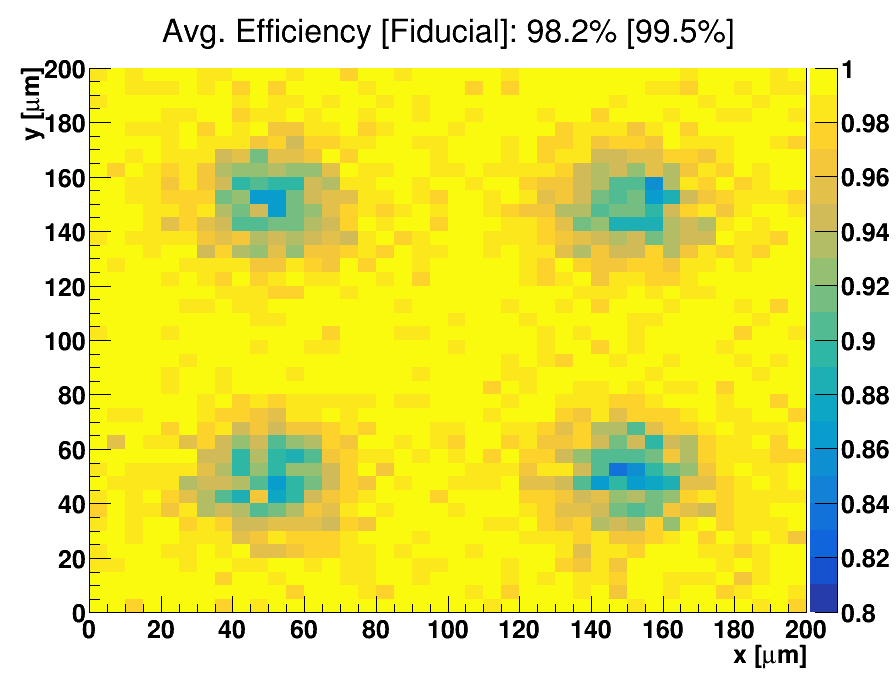}
		\caption{$\bias=\SI{400}{\volt}$}
	\end{subfigure}
	\begin{subfigure}[t]{.48\textwidth}
		\includegraphics[width=\linewidth]{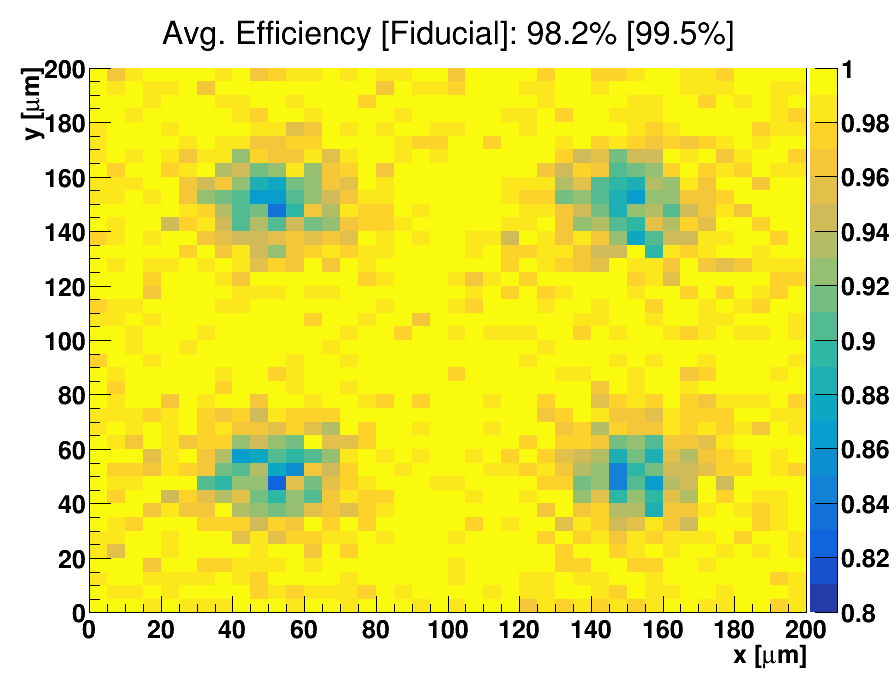}
		\caption{$\bias=\SI{500}{\volt}$}
	\end{subfigure} \hfill%
	\begin{subfigure}[t]{.48\textwidth}
		\includegraphics[width=\linewidth]{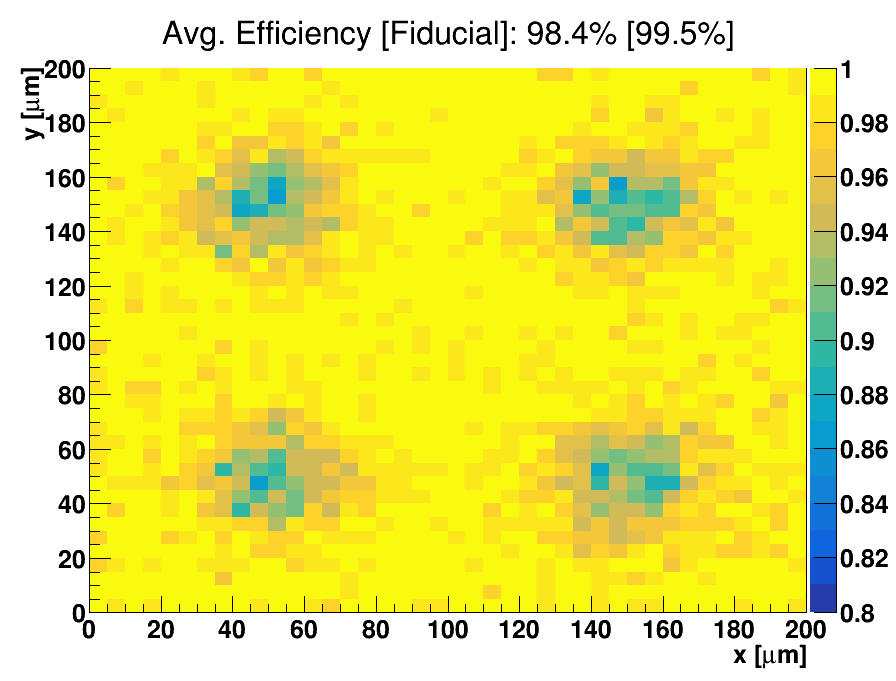}
		\caption{$\bias=\SI{600}{\volt}$}
	\end{subfigure}
	\caption{In-pixel efficiency maps, as defined in Section~\ref{sec:tbmon2}, for device DUT-B (linear FE) with a threshold of \SI{1200}{\electron}.}
	\label{fig:results:Dec18 in-pixel eff thl1200}
\end{figure}

\begin{figure}[!htb]
	\centering
	\begin{subfigure}[t]{.48\textwidth}
		\includegraphics[width=\linewidth]{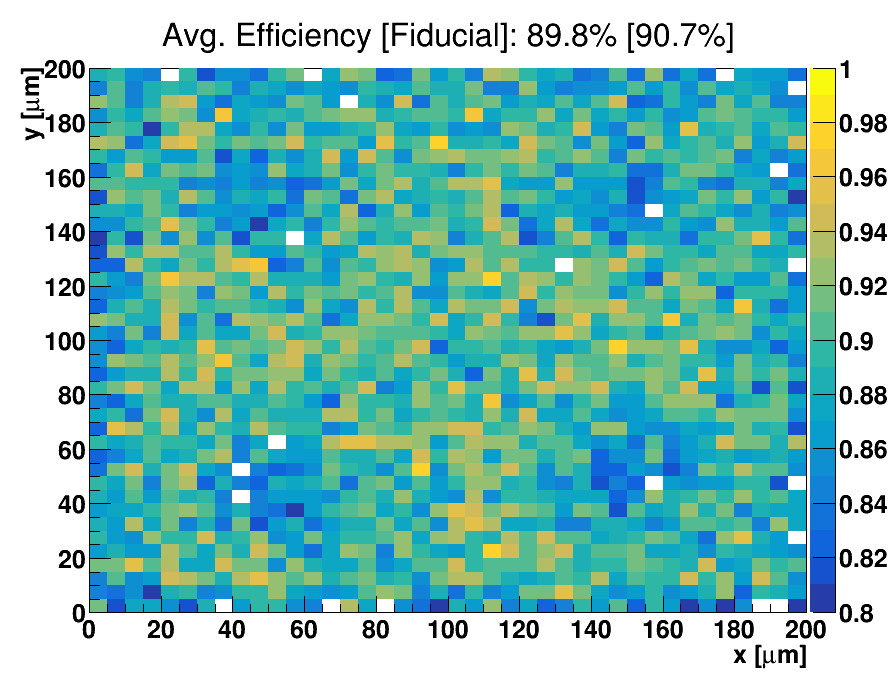}
		\caption{$\text{threshold}=\SI{900}{\electron}$}
	\end{subfigure} 
	\begin{subfigure}[t]{.48\textwidth}
		\includegraphics[width=\linewidth]{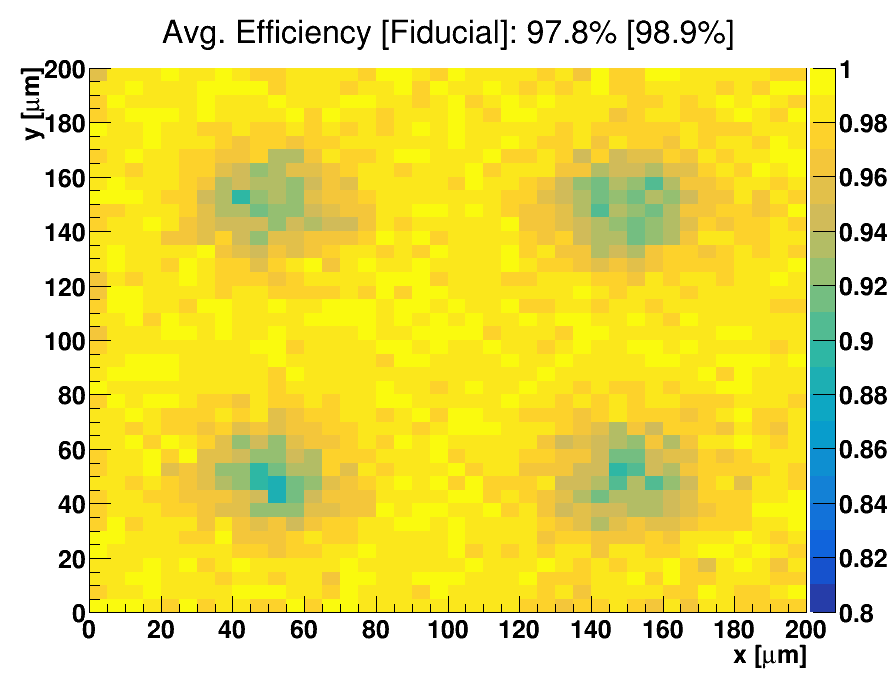}
		\caption{$\text{threshold}=\SI{1000}{\electron}$}
	\end{subfigure} 
	\begin{subfigure}[t]{.48\textwidth}
		\includegraphics[width=\linewidth]{figures/results/december18/tbmon/09Lin_V600_th1200.png}
		\caption{$\text{threshold}=\SI{1200}{\electron}$}
	\end{subfigure} 
	\caption{In-pixel efficiency maps, as defined in Section~\ref{sec:tbmon2}, for device DUT-B~(linear FE) with $\bias=\SI{600}{\volt}$}
	\label{fig:results:Dec18 in-pixel eff V600}
\end{figure}

\begin{figure}[!htb]
	\centering
	\begin{subfigure}[t]{.48\textwidth}
		\includegraphics[width=\linewidth]{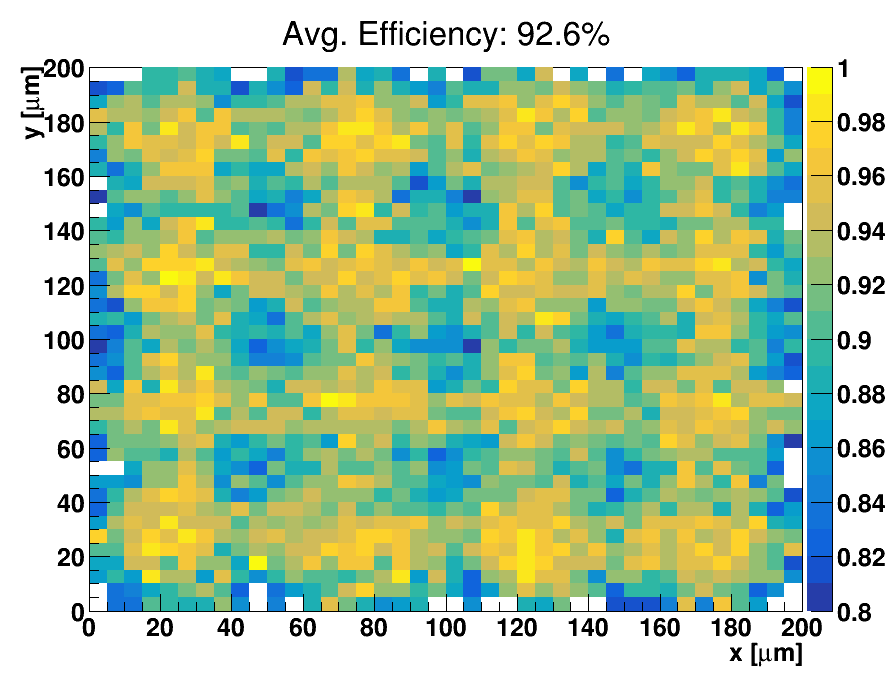}
		\caption{$\bias=\SI{200}{\volt}$}
	\end{subfigure} \hfill%
	\begin{subfigure}[t]{.48\textwidth}
		\includegraphics[width=\linewidth]{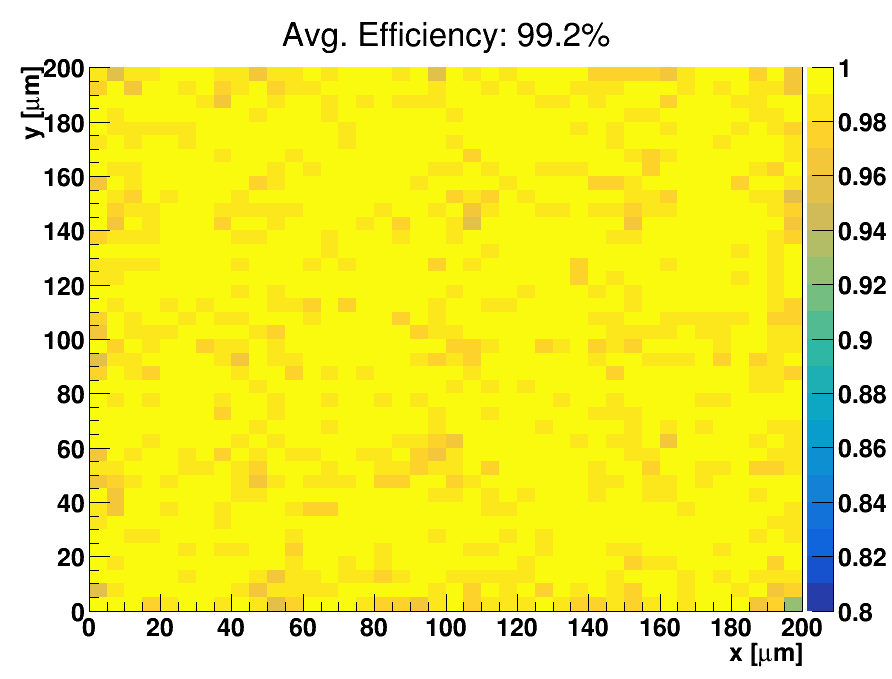}
		\caption{$\bias=\SI{400}{\volt}$}
	\end{subfigure}
	\begin{subfigure}[t]{.48\textwidth}
		\includegraphics[width=\linewidth]{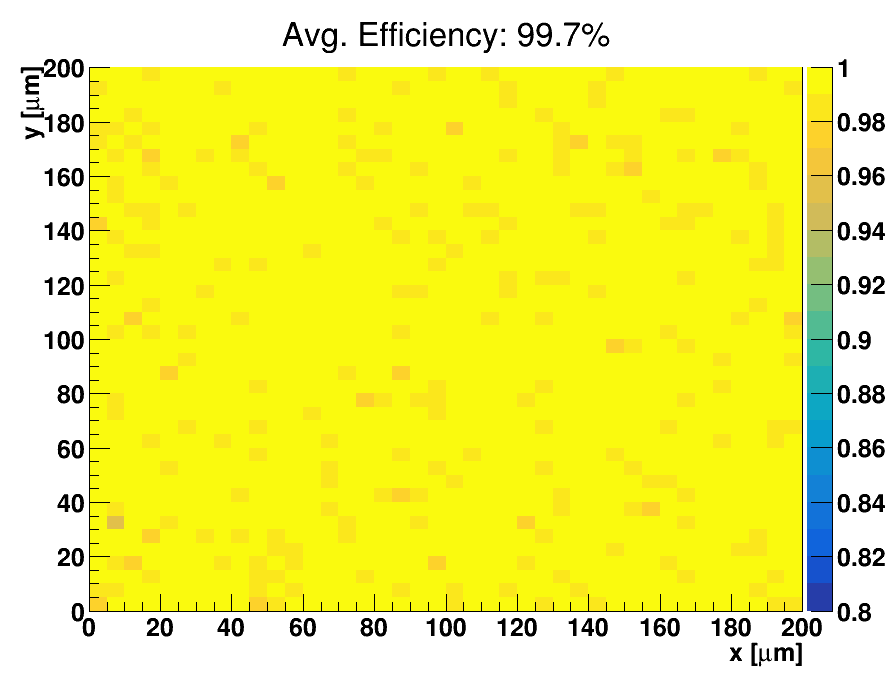}
		\caption{$\bias=\SI{600}{\volt}$}
	\end{subfigure} \hfill%
	\caption{In-pixel efficiency maps, as defined in Section~\ref{sec:tbmon2}, for device DUT-A (linear FE) with $\text{threshold}=\SI{1027}{\electron}$.}
	\label{fig:results:Dec18 in-pixel eff 45 thl 1027}
\end{figure}

Figure~\ref{fig:results:Dec18 in-pixel eff thl1200} shows the in-pixel efficiency maps
for device DUT-B at different $\bias$ points, with a threshold of \SI{1200}{\electron}. Here, the efficiency of the device increases as \bias is increased. Despite reaching near-perfect efficiency across the majority of the sensor, the regions around the punch-through dots retain their lower efficiency.
Figure~\ref{fig:results:Dec18 in-pixel eff V600} shows the in-pixel efficiency maps
for the same device at different threshold points and $\bias=\SI{600}{\volt}$. Here, the efficiency of the device increases as the threshold is increased, until the efficiency reaches a plateau at the higher threshold values. Beyond this, an increase in the threshold begins to remove signal hits, to the detriment of the device efficiency. As with the \bias scan, the regions around the punch-through dots have a persistently lower efficiency than the other areas on the sensor, which reach near-perfect efficiency.
Figure~\ref{fig:results:Dec18 in-pixel eff 45 thl 1027} shows the in-pixel efficiency maps
for device DUT-A at different $\bias$ points, with a threshold of \SI{1027}{\electron}. Here, the efficiency of the device increases as the threshold is increased. Despite reaching near-perfect efficiency across the majority of the sensor at higher threshold values, the regions around the pixel corners exhibit lower efficiency at lower threshold as a result of charge-sharing.

\clearpage

\subsection{Cross-testbeam comparisons}\label{sec:crossTB}

Comparisons are performed between the October 2018 and December 2018 testbeams where the same devices were tested during both campaigns. Where overlapping operating parameters are present, bias and threshold scans are combined and summarised.

\begin{figure}[!htb]
	\centering
	\includegraphics[width=\linewidth]{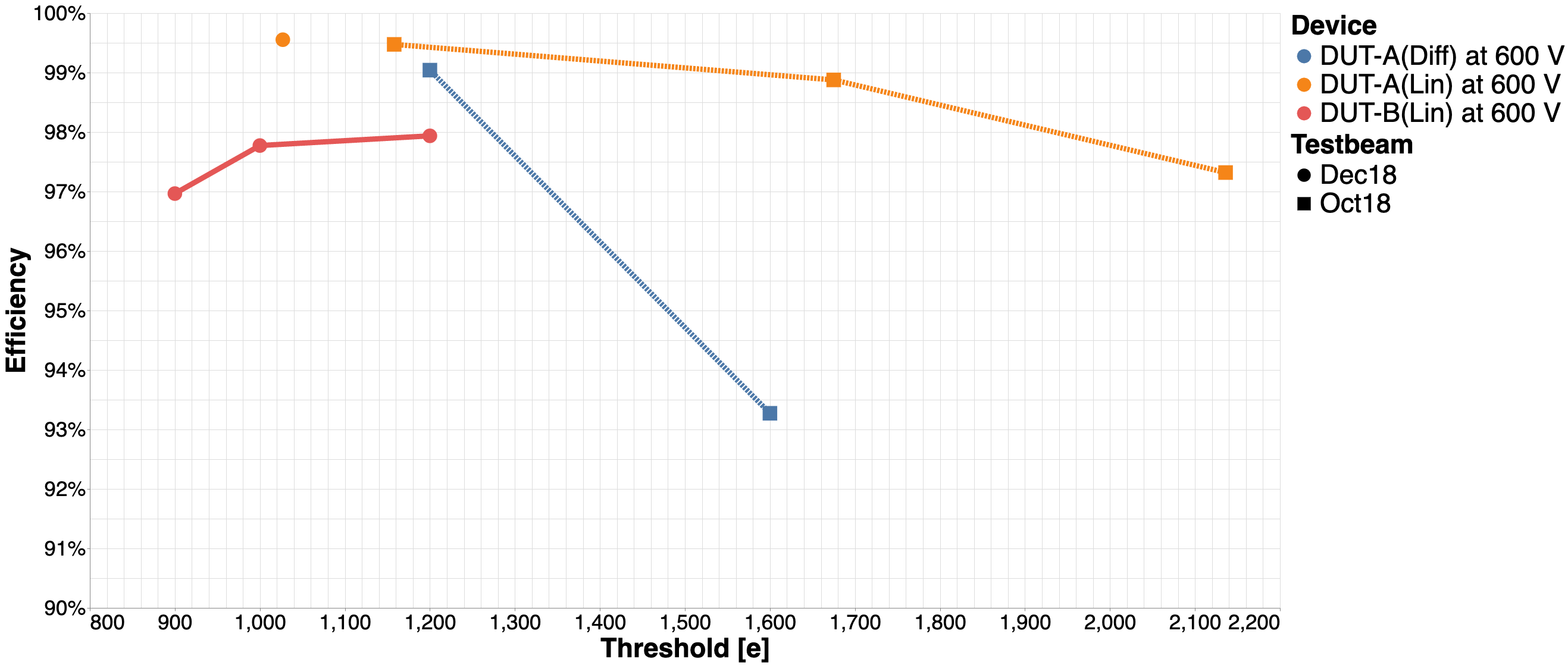}
	\caption{Overal pixel efficiency as a function of threshold for both October and December testbeam campaigns, selecting runs with $\bias=\SI{600}{\volt}$.}
	\label{fig:cross_oct_dec_600V}
\end{figure}

Figure~\ref{fig:cross_oct_dec_600V} shows the overall efficiency as a function of the threshold for
all devices under test at both October and December testbeam campaigns,
selecting runs with $\bias=\SI{600}{\volt}$.
The efficiency generally increases when increasing the threshold to above \SI{1000}{\electron}, at which point the effect of noise is reduced.
Once the threshold reaches an optimal working point, efficiency will decrease as the threshold is increased further, as signal loss begins to occur.
For all devices under test, the desired $97\%$ efficiency is surpassed for optimal operation thresholds.

\begin{figure}[!htb]
	\centering
	\includegraphics[width=\linewidth]{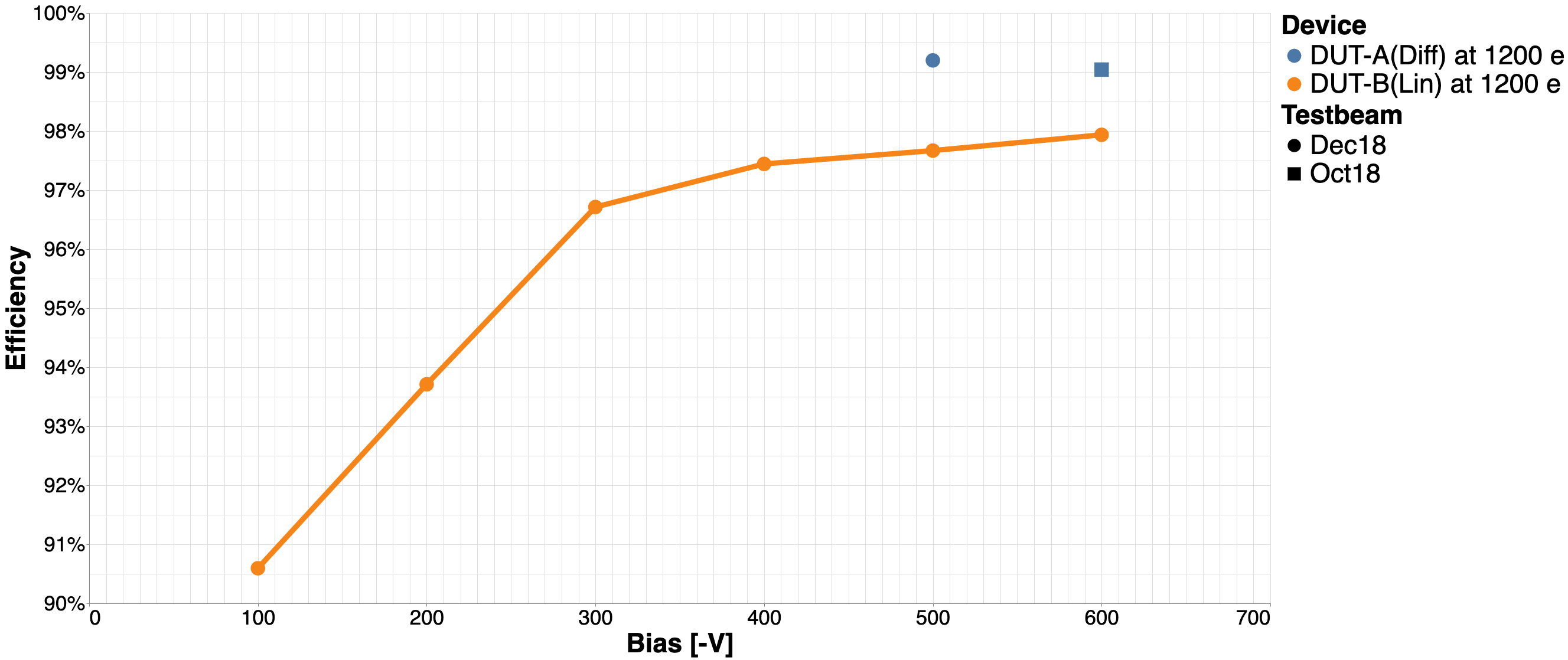}
	\caption{Overall efficiency as a function of $\bias$ comparing data
	from both testbeam campaigns, selecting the runs with FE threshold of $\text{th}=\SI{1200}{\electron}$.}
\label{fig:cross_oct_dec_1200e}
\end{figure}

Figure~\ref{fig:cross_oct_dec_1200e} shows the overall efficiency as a function of $\bias$
for both testbeam campaigns, selecting runs with FE threshold of $\text{th}=\SI{1200}{\electron}$.
As discussed in Section~\ref{sec:results_dec2018}, a turn-on curve is observed for device DUT-B~(lin).
The required efficiency is observed for all devices with $\bias\ge\SI{400}{\volt}$. 

\section{Conclusions}\label{sec:conclusions}
Hybrid pixel detectors consisting of a planar silicon sensor bump-bonded to an RD53A readout chip have been characterised by the analysis of reconstructed testbeam data. Measurements were made using an EUDET-like beam telescope over two testbeam campaigns at the CERN SPS and DESY testbeam facilities in October 2018 and December 2018, respectively. Two devices were measured, with and without punch-through biasing structures, and using both linear and differential readout modes. Both devices used sensors with pixel pitch~$50\times50$~\SI{}{\um\squared}, thickness of \SI{150}{\um} and irradiated to $3.4\times10^{15}$~\SI{}{\nequ\per\cm\squared}.
	
A range of \bias and threshold scans were performed on the devices, with each device required to surpass 97\% efficiency at $\bias=\SI{400}{\volt}$.

Across devices, the efficiency is observed to increase as the \bias or threshold is increased until reaching an efficiency plateau. When the threshold is increased further, it is seen to reduce the device efficiency as signal hits begin to be cut away.

Devices which feature a punch-though bias structure are seen to reach an efficiency plateau at lower \bias and threshold than those with no such structure. The efficiency at the plateau is also observed to be lower than the device with no biasing structure. In-pixel efficiency maps show regions of low efficiency around the punch-through dots on the device with the biasing structure. 

At $\bias=\SI{600}{\volt}$, the highest global efficiency was observed on device \mbox{DUT-A}, using the linear front-end with no punch-through biasing structure, at  99.6\% efficiency, at a threshold of $\text{threshold}=\SI{1158}{\electron}$. At the same \bias, threshold, and operating mode, device \mbox{DUT-B} with a \mbox{punch-through} bias structure was observed to have a global efficiency of 98.4\%  and an efficiency of 99.5\% in the fiducial area which masks out the regions contaminated by the biasing structure.

Although the biasing structure is seen to have a detrimental effect on the efficiency of the sensor, devices both with and without this feature comfortably surpass the 97\% efficiency criteria. As such, the benefits provided by these biasing structures for production of modules is considered to outweigh the slight drop in efficiency which does not impact the ability of the device to pass QC criteria.

Device DUT-A was observed to achieve greater efficiency using the linear front-end than when using the differential front-end. At $\bias=\SI{600}{\volt}$ and a threshold of around \SI{1200}{\electron}, the linear front-end produced a global efficiency of 99.6\% whilst the differential front-end produced a global efficiency of 99.2\%. In addition to this, the efficiency of the differential front-end was observed to fall off more quickly as the threshold was raised. At $\bias=\SI{600}{\volt}$ and a threshold of around \SI{1600}{\electron}, the linear front-end produced a global efficiency of 98.9\% whilst the differential front-end produced a global efficiency of 93.6\%. In the absence of thresholds below around \SI{1200}{\electron} to make comparisons with, and given that the differential front-end is expected to be capable of operating at lower threshold than the linear front-end, it is possible that the efficiency of the differential front-end simply peaks at lower threshold than the linear front-end and could therefore outperform the linear front-end at these lower thresholds. Additional data points at these lower thresholds would be required to verify this.

\newpage
\acknowledgments

The measurements leading to these results were made at the testbeam facility at DESY Hamburg (Germany),
a member of the Helmholtz Association (HGF). We gratefully thank the operators of this facility.

The reconstruction and analysis presented here was performed within a framework based on one originally developed in collaboration with Tobias Fitschen. Our thanks to him for this.

\newpage
\bibliographystyle{latex/JHEP}
\bibliography{aPaper}

\providecommand{\href}[2]{#2}\begingroup\raggedright\begin{thebibliography}{10}

\bibitem{macchiolo_phase-2_2020}
A.~Macchiolo, \emph{The {Phase}-2 {ATLAS} {ITk} pixel upgrade},
  \href{https://doi.org/10.1016/j.nima.2019.06.002}{\emph{Nuclear Instruments
  and Methods in Physics Research Section A: Accelerators, Spectrometers,
  Detectors and Associated Equipment} {\bfseries 962} (2020) 162261}.

\bibitem{keller_atlas_2020}
J.~S. Keller, \emph{The {ATLAS} {ITk} strip detector system for the {High}
  {Luminosity} {LHC} upgrade},
  \href{https://doi.org/10.1016/j.nima.2019.04.007}{\emph{Nuclear Instruments
  and Methods in Physics Research Section A: Accelerators, Spectrometers,
  Detectors and Associated Equipment} {\bfseries 958} (2020) 162053}.

\bibitem{M_bius_2022}
S.~Möbius, \emph{Module development for the {ATLAS} {ITk} pixel detector},
  \href{https://doi.org/10.1088/1748-0221/17/03/c03042}{\emph{Journal of
  Instrumentation} {\bfseries 17} (2022) C03042}.

\bibitem{Garcia-Sciveres:2287593}
{\scshape RD53 Collaboration} collaboration, \emph{{The RD53A Integrated
  Circuit}},  tech. rep., CERN, Geneva, Oct, 2017.
\newblock
  \href{https://cds.cern.ch/record/2287593}{https://cds.cern.ch/record/2287593}.

\bibitem{KIT}
{Karlsruhe Institute of Technology}, ``{KIT Proton Irradiation}.''
  \url{https://www.etp.kit.edu/english/264.php}.

\bibitem{ATLAS-TDR-30}
{ATLAS Collaboration}, \emph{{ATLAS Inner Tracker Pixel Detector: Technical
  Design Report}},  tech. rep., 2017.
\newblock
  \href{https://cds.cern.ch/record/2285585}{https://cds.cern.ch/record/2285585}.

\bibitem{Wenninger_2021}
J.~Wenninger, ``{SPS} operation.'' \url{https://jwenning.web.cern.ch/SPS.html}.

\bibitem{Jansen_2016}
H.~Jansen et~al., \emph{{Performance of the EUDET-type beam telescopes}},
  \href{https://doi.org/10.1140/epjti/s40485-016-0033-2}{\emph{EPJ Tech.
  Instrum.} {\bfseries 3} (2016) 7}
  [\href{https://arxiv.org/abs/1603.09669}{{\ttfamily 1603.09669}}].

\bibitem{Daas_2021}
M.~Daas, Y.~Dieter, J.~Dingfelder, M.~Frohne, G.~Giakoustidis, T.~Hemperek
  et~al., \emph{{BDAQ}53, a versatile pixel detector readout and test system
  for the {ATLAS} and {CMS} {HL}-{LHC} upgrades},
  \href{https://doi.org/10.1016/j.nima.2020.164721}{\emph{Nuclear Instruments
  and Methods in Physics Research Section A: Accelerators, Spectrometers,
  Detectors and Associated Equipment} {\bfseries 986} (2021) 164721}.

\bibitem{DESY_2021}
{DESY}, ``Generation of the desy test beams.''
  \url{https://particle-physics.desy.de/test_beams_at_desy/e252106/e252211}.

\bibitem{Himmi_2008}
A.~Himmi et~al., \emph{{Mimosa26 User Manual}},  tech. rep., December, 2008.
\newblock
  \url{https://www-rnc.lbl.gov/hft/hardware/docs/Phase1/M26_UserManual.pdf}.

\bibitem{RD53A_manual}
{\scshape RD53} collaboration, \emph{{The RD53A Integrated Circuit}},  tech.
  rep., CERN, Geneva, Oct, 2017.
\newblock
  \href{https://cds.cern.ch/record/2287593}{https://cds.cern.ch/record/2287593}.

\bibitem{FEI4_manual}
M.~Garcia-Sciveres et~al., \emph{{The FE-I4 pixel readout integrated circuit}},
  \href{https://doi.org/10.1016/j.nima.2010.04.101}{\emph{Nucl. Instrum. Meth.
  A} {\bfseries 636} (2011) S155}.

\bibitem{Baesso_2019}
P.~Baesso, D.~Cussans and J.~Goldstein, \emph{{The AIDA-2020 TLU: a flexible
  trigger logic unit for test beam facilities}},
  \href{https://doi.org/10.1088/1748-0221/14/09/P09019}{\emph{JINST} {\bfseries
  14} (2019) P09019} [\href{https://arxiv.org/abs/2005.00310}{{\ttfamily
  2005.00310}}].

\bibitem{Bisanz_2020}
T.~Bisanz, H.~Jansen, J.-H. Arling, A.~Bulgheroni, J.~Dreyling-Eschweiler,
  T.~Eichhorn et~al., \emph{Eutelescope: A modular reconstruction framework for
  beam telescope data},
  \href{https://doi.org/10.1088/1748-0221/15/09/p09020}{\emph{Journal of
  Instrumentation} {\bfseries 15} (2020) P09020–P09020}.

\bibitem{Gaede_2003}
F.~Gaede, T.~Behnke, N.~Graf and T.~Johnson, \emph{{LCIO: A Persistency
  framework for linear collider simulation studies}}, {\emph{eConf} {\bfseries
  C0303241} (2003) TUKT001}
  [\href{https://arxiv.org/abs/physics/0306114}{{\ttfamily physics/0306114}}].

\bibitem{GEAR}
{ilcSoft community}, ``{GEometry API for Reconstruction}.''
  \url{https://ilcsoft.desy.de/portal/software_packages/gear/}.

\bibitem{Ahlburg_2020}
P.~Ahlburg, S.~Arfaoui, J.-H. Arling, H.~Augustin, D.~Barney, M.~Benoit et~al.,
  \emph{Eudaq—a data acquisition software framework for common beam
  telescopes},
  \href{https://doi.org/10.1088/1748-0221/15/01/p01038}{\emph{Journal of
  Instrumentation} {\bfseries 15} (2020) P01038–P01038}.

\bibitem{Kleinwort_2012}
C.~Kleinwort, \emph{General broken lines as advanced track fitting method},
  \href{https://doi.org/10.1016/j.nima.2012.01.024}{\emph{Nuclear Instruments
  and Methods in Physics Research Section A: Accelerators, Spectrometers,
  Detectors and Associated Equipment} {\bfseries 673} (2012) 107–110}.

\bibitem{tbmon2}
tbmon2 community, ``{TBmon2 - Testbeam Data Analysis Software}.''
  \url{https://gitlab.cern.ch/tbmon2/tbmon2}.

\end{thebibliography}\endgroup
\end{document}